\documentclass{article}

\usepackage{arxiv}

\usepackage[utf8]{inputenc} 
\usepackage[T1]{fontenc}    
\usepackage{hyperref}       
\usepackage{url}            
\usepackage{booktabs}       
\usepackage{amsfonts}       
\usepackage{nicefrac}       
\usepackage{microtype}      
\usepackage{lipsum}
\usepackage[]{algorithm2e}
\usepackage{lipsum,graphicx,subcaption}
\usepackage{amssymb}
\usepackage{float}
\usepackage{amsmath}
\usepackage{tabularx}
\usepackage{gensymb}

\title{Identifying Vulnerabilities of Industrial Control Systems using Evolutionary Multiobjective Optimisation}

\author{Nilufer Tuptuk \\
  Department of Computer Science\\
  University College London\\
  London, United Kingdom \\
  \texttt{n.tuptuk@ucl.ac.uk} \\
   \And
Stephen Hailes \\
  Department of Computer Science\\
  University College London\\
  London, United Kingdom \\
  \texttt{s.hailes@ucl.ac.uk} \\
}

\begin{document}
\maketitle

\begin{abstract}

In this paper we propose a novel methodology to assist in identifying vulnerabilities in a real-world complex heterogeneous industrial control systems (ICS) using two evolutionary multiobjective optimisation (EMO) algorithms, NSGA-II and SPEA2. Our approach is evaluated on a well known benchmark chemical plant simulator, the Tennessee Eastman (TE) process model. We identified vulnerabilities in individual components of the TE model and then made use of these to generate combinatorial attacks to damage the safety of the system, and to cause economic loss. Results were compared against random attacks, and the performance of the EMO algorithms were evaluated using hypervolume, spread and inverted generational distance (IGD) metrics. A defence against these attacks in the form of a novel intrusion detection system was developed, using a number of machine learning algorithms. Designed approach was further tested against the developed detection methods. Results demonstrate that EMO algorithms are a promising tool in the identification of the most vulnerable components of ICS, and weaknesses of any existing detection systems in place to protect the system. The proposed  approach can be used by control and security engineers to design security aware control, and test the effectiveness of security mechanisms, both during design, and later during system operation.
\end{abstract}


\section{Introduction}

Industrial Control Systems (ICS) are command and control systems that are found at the core of the national critical infrastructure services such as gas; electricity; oil; water supply; telecommunication; transportation; process manufacturing (chemicals, pharmaceuticals, paper, food, beverages and other batched-based manufacturers); and discrete manufacturing (automobiles, ships, computers and many other durable goods). The security of ICS is of critical importance in industrialised economies: they are so pervasive that national security, public health and safety, and economic growth all rely on their correct operation. In the past, security of ICS was achieved simply through isolation and control of physical access. However, ICS are making increasing use of network technologies, commercial-off-the-shelf (COTS) components, and wireless systems because of factors such as low-cost, increased sensing and communication capacity and convenience.

With these technological advances, factory and plant networks are becoming highly connected over multiple layers, and signals sent between control components (i.e. sensors, controllers, and actuators), are sent through a common network, known as networked control system \cite{PATTON2007280}. The number of motivated and highly skilled adversaries carrying out complex attacks against networked control system is on the increase. Some of the past attacks include the attack against the operational systems of Evraz Steel in North America \cite{evraz}; attack on Ukraine's power grid \cite{crashoverride} that targeted the electric transmission system in Kiev; attack against a German steel mill \cite{german:steel} that caused  unspecified but ``massive" physical damage; malware attacks such as Duqu \cite{duqu} and Havex \cite{havex} that targetted ICS for industrial espionage; and Stuxnet \cite{Stuxnet} that targetted Iran's Natanz nuclear plant, and destroyed centrifuges installed at the time of the attack. As the evidence from these attacks show the potential outcome of a successful attack on a critical service ranges from injuries and fatalities, through serious damage to the environment, to catastrophic nation-wide economic loss due to production losses or degradation of products and services. Shutting down or preventing access to these systems, even for a short time, may cause significant harm to people and may impact public confidence, leading to a general feeling of insecurity. 

Despite technological advances in ICS, understanding the vulnerabilities of ICS at the process level and the potential physical consequences when these vulnerabilities are exploited remains an important aspect of ICS security that is currently lacking research. It is critical to national economic resilience that we explore better ways of searching for vulnerabilities in the industrial processes and defence mechanisms, and understand the impact of these vulnerabilities would be if they were to be exploited. Considerable research \cite{10.1145/1966913.1966959, HUANG200973, Krotofil2013, Genge12impactof, WANG201824, 10.1007/978-3-642-45330-4_15, 8270567} has focused on developing threat models and analysing a variety of attacks, aimed at modifying process measurements (sensor measurements) or manipulated variables (values going from controller to actuators), or manipulating the control algorithm (e.g. set points), however there is a lack of research work on investigating impact of combinatorial attacks, and automating attack generation.

In this paper, we seek to identify the most vulnerable components in the ICS by searching for process level attacks that cause the greatest damage in terms of plant safety and economics, using the least effort, as well as identifying weaknesses in detection methods, at the lowest risk of being caught. To solve this, we establish the problem as a multiobjective optimisation problem, and investigated effectiveness of EMO algorithms as a tool for generating attacks.

Multiobjective optimisation has been widely applied to real world complex scientific and engineering problems, but, we are not aware of any studies using EMO algorithms to identify the most vulnerable components of cyber-physical systems, including ICS. However, evolutionary algorithms have been extensively used to improve the security mechanisms used especially within IT networks in a wide variety of ways. Others have used genetic algorithm (GA) \cite{Li04usinggenetic, 1460505, 5949394, gaids, DBLP:journals/corr/abs-1204-1336, Diaz-gomez05improvedoff-line, Goyal2008}, and genetic programming (GP) \cite{COIN:COIN247, 5718700} to evolve new rules to detect new forms of network intrusion. Our previous work \cite{Mrugala:2016:GECCOcomp} \cite{7979715} used GP to attack a wireless sensor network (WSN) protected by an artificial immune intrusion detection system. Kayacik et al. \cite{Kayacik:2006:EBO:1143997.1144271} used GP to evolved variants of buffer overflow attacks against an open-source signature-based IDS. John et al. \cite{John:2014:EBM:2598394.2605437} applied GA to the improvement of moving target defence, in which the defence changes the system’s attack surface to disrupt the intelligence gathered by the attacker. Dewri et al. \cite{Dewri:2007:OSH:1315245.1315272} used GA with multiobjective optimisation to investigate optimal security measures for a system.  Garcie et al. \cite{Garcia:2017:ICA:3067695.3076081}, Hemberg et al. \cite{10.1145/3205651.3208287}, Rush et al. \cite{Rush:2015:CAN:2739482.2768429} used co-evolution to model attacker and defence dynamics for network security. Co-evolutionary concepts have also been investigated to prevent faults and cascading blackouts in electric power transmission systems \cite {4290990, doi:10.1142/S1793005709001416}; automate red teaming for military scenarios \cite {5679047}; and improve the performance of malware detectors \cite{8506545}. 

Our work complements the existing studies in applying evolutionary multiobjective optimisation to identify ICS vulnerabilities by generating attacks. We investigate both the worst-case condition where ICS has no security protection, and also where there are some measures against attacks, in the form of a novel intrusion detection system.  

The remainder of the paper is organised as follows. Section 2 presents the  background material related to our approach. This includes the characteristics of the the case study, the Tennessee Eastman (TE) process; description of the multiobjective optimisation problems; and attack model used for generating attacks. Section 3 covers the baselines to compare EMO approach; details of the EMO algorithms; the detection methods used to evolve attacks against detection; the performance metrics used for comparing EMO algorithms; and describes the random approach used to compare EMO approach. Section 4 presents experimental results. In section 5, application of results are discussed. Conclusion and future work are presented in Section 6. 

\section{Background}

\subsection{Case study: The Tennessee Eastman (TE) process }
\label{casestudy}
To develop and explore the effectiveness of evolutionary multiobjective optimisation as an possible candidate for identifying vulnerabilities in ICS, a well-known chemical plant, the Tennessee Eastman (TE) process control model \cite{DOWNS1993245} was selected. 

\begin{figure*}[h!]
\begin{centering}
\includegraphics[width=170mm]{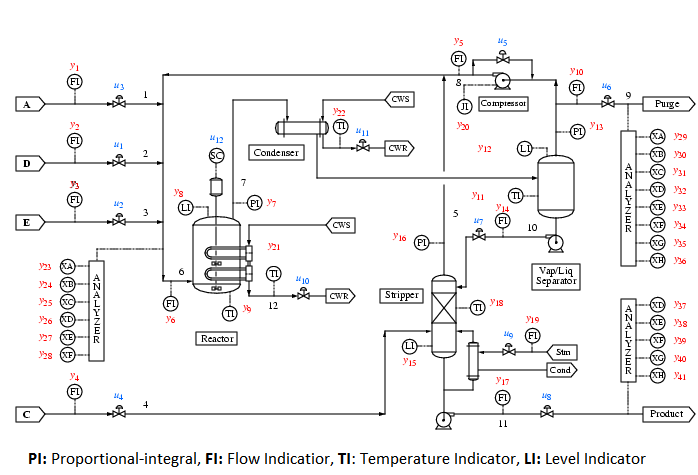}
\caption{Tennessee Eastman challenge model \cite{DOWNS1993245, ie000586y}}
\label{temodel}
\end{centering}
\end{figure*}

Our reasons for selecting this process are: i) it is a well-known model that has been widely studied; ii) it is a complex, highly non-linear system with a number of components that reflects a real process; iii) safety and economic viability can be quantified; iv) the code and model is available, and have been revised and validated over the years; v) it continues to be a relevant model for both the control, and more recently, the security communities. We are not aware of any other open source model that has these properties. The TE model is based on a real chemical process; however, the identity of the reactants and products are hidden to maintain commercial confidentiality. The process has eight components: four gaseous reactants {(A, C, D, E)}, two products {(G, H)}, an inert component {(B)} and a by-product {(F)}. These reactions are \cite{DOWNS1993245}: 

	\begin{flushleft}
	{A(g)} {+} {C(g)} {+} {D (g)} $\rightarrow G(liq)$, Product 1, \newline
	{A(g)} {+} {C (g)} {+} {E (g)} $\rightarrow H(liq)$, Product 2,  \newline
	{A(g)} {+} {E (g)}  $\rightarrow  ${F(liq), Byproduct},  \newline
  {3D(g)} $\rightarrow {2F(liq)}$, Byproduct.  \newline
\end{flushleft}

The process is illustrated in Figure \ref{temodel}, and consists of five major components \cite{DOWNS1993245}: a reactor, a product condenser, a vapour-liquid separator, a recycle compressor and a product stripper. 

There are 41 process measurement variables known as XMEASs (sensors, denoted as y signals in red) and 12 manipulated variables known as XMVs (valves/actuators, denoted as u signals in blue), illustrated in Figure \ref{temodel}, that are involved in controlling and monitoring the plant. The main control objectives of the plant are \cite{DOWNS1993245}: to maintain the process variables at the desired values; to ensure that the operational conditions are within the equipment constraints; and to minimise variability of the product rate and product quality during disturbances. To protect the safety of the process, the TE model has a set of operating constraints that are captured as normal operating limits and shutdown operating limits. If the process reaches the shutdown limits, it automatically shuts the plant down. These constraints (low/high limits of reactor pressure, reactor level, reactor temperature, product separator level and stripper base level) \cite{DOWNS1993245} are put in place to protect the personnel, equipment, production, and meet compliance requirements. The operating costs of the TE process are calculated according to the following equation \cite{DOWNS1993245}: 
	\begin{flushleft}
total costs = (purge costs)(purge rate)+(product stream costs)(product rate) + 
(compressor costs)(compressor work)+(steam costs)(steam rate) 
\end{flushleft}

The TE process problem makes no recommendation as to what needs to be controlled, and leaves the selection of controlled variables and control strategies to the control engineers. Most proposed solutions do not control all the variables. The control strategy used in this paper is that described by Larsson et al. in \cite{ie000586y} using 16 process measurements and 9 manipulated variables. The process model used in this paper is developed by Ricker, available from his home page \cite{Ricker15}. The code is implemented in C, with a MATLAB/Simulink interface via an S-function implementation. Isakov and Krotofil \cite{IsakovKrotofil} extended the original Simulink model by enhancing it with Simulink blocks that enable integrity and denial-of-service (DoS) attacks to be carried out on the sensors and manipulated variables. We extended their model with replay attacks, and made further small changes needed to carry out the work in this paper.

We first analyse the vulnerabilities of the system by considering two types of adversary: i) an adversary targeting the safety of the plant by attempting to shut it down using the least effort; ii) an adversary targeting the operating cost of the plant to increase economic loss using the least effort. The effort of the attack is calculated as the number of sensors and actuators that must be compromised. 

We, then, use machine learning techniques to detect attacks, and evolve new attacks against the detection methods. We assume an adversary who has control over her attack vector and who knows the feature space used by the detection system. However, she does not know the details of the underlying detection methods. So, essentially, the detection is a blackbox that can be queried. An adversary queries the detection with the attack vector, and obtains a detection probability. Her goal is to find attacks that cause damage whilst evading detection, and using the least effort. Thus, the problem is formulated as a search problem with objectives. 

\subsection{Multiobjective optimisation}
In many real-world problems, decisions need to be made on the basis of multiple competing or conflicting objectives and constraints. This is often the case when making decisions related to cyber security investment or security hardening, attempting to balance risks of attack against a limited budget to buy defensive countermeasures. In such situations, formulating the problem as a multiobjective optimisation (MOO) with multiple choices can help to determine the trade-offs among the objectives in a more effective manner \cite{7360024, FIELDER201613, Dewri:2007:OSH:1315245.1315272}. These approaches search for the set of {\em non-dominated} or \textit{Pareto-optimal} solutions. A solution is defined non-dominated if there are no other solutions that would improve any objective without degrading one or more of the other objectives. Once the set of Pareto-optimal solutions, has been identified, a decision maker can make a decision by examining the tradeoffs represented by individual solutions within the set. MOO takes a problem with multiple objectives and simultaneously seeks to optimise all objectives, providing solutions in, or close to, the true Pareto-optimal set. More often than not, this is an estimate because determining the true Pareto-optimal set for real world problems is hard, either because the search space is too large or because obtaining solutions is costly in time and computation. 

In this piece of work, we are concerned with finding optimised solutions for the following multiobjective problems:
\begin{enumerate}
    \item \textbf{Attack safety of the plant (shutting down the plant)}: Minimise plant operating time of the plant, minimise effort required to carry out attacks.
    \item \textbf{Increase operating cost (economic loss attack):} Maximise operating cost of the plant, minimise effort required to carry out attacks.
    \item \textbf{Increase economic loss and avoid detection:} Maximise operating cost of the plant, minimise detection (alarm) probability, and minimise effort required to carry out attacks.
\end{enumerate}
The generation of optimal attacks against real systems with large number of components involves selection of many parameters: attack targets (controllers, sensors, actuators); attack types; and attack parameters for these attacks (e.g. mode of attacks, attack start times and attack duration). As explained earlier, in practice identifying the true Pareto-optimal set to such problems may not be feasible. Evolutionary multiobjective optimisation is a promising approach to identify an estimate of best trade-off attacks in such complex systems at reasonable computational cost.

\subsection{Attack modelling}

\begin{figure} 
\centering
      \includegraphics[width=10cm]{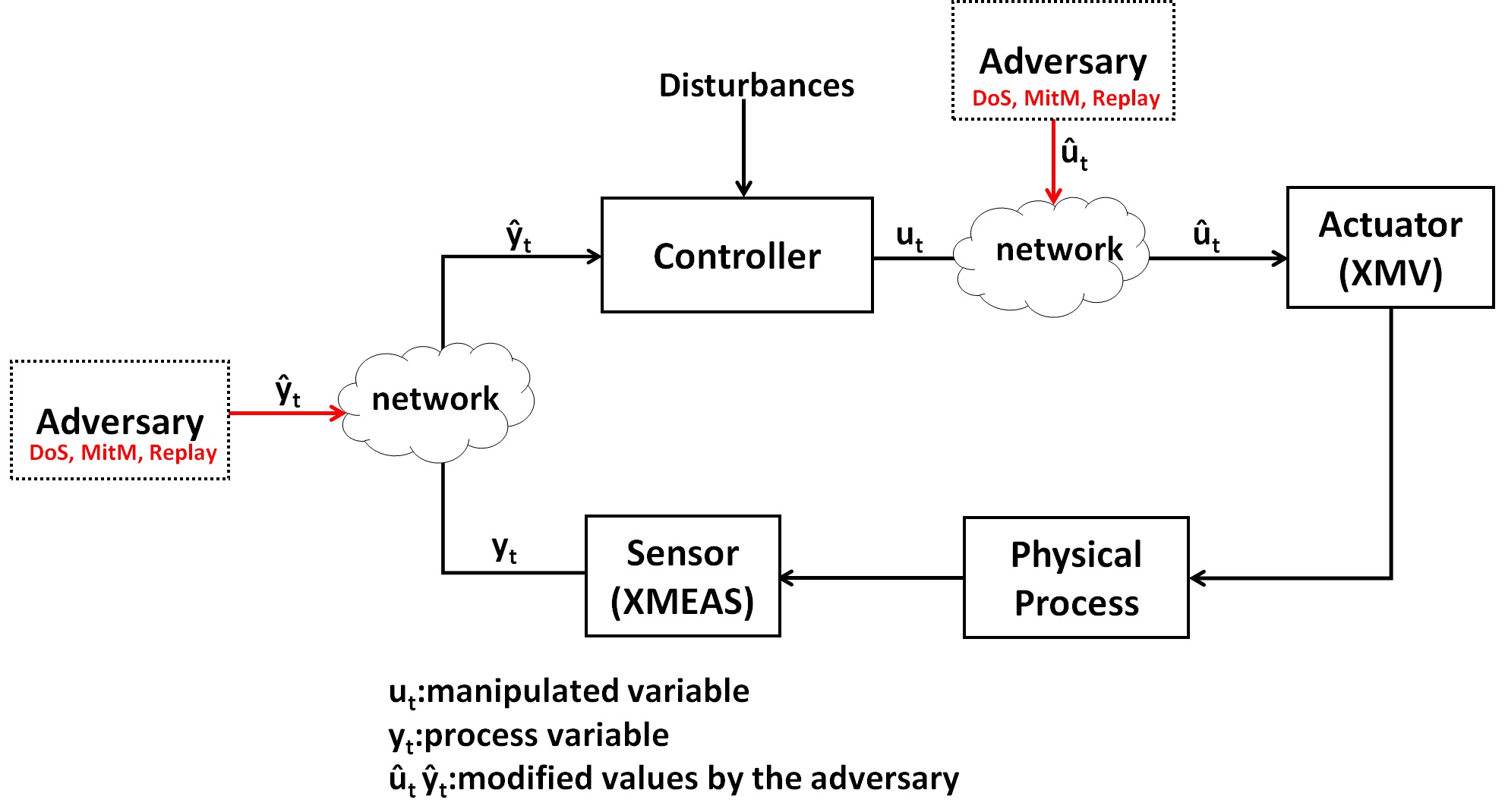}
    \caption{ICS attack model against the networked control system}
    \label{adversarythreatmodel}
\end{figure}

Figure \ref{adversarythreatmodel} shows our underlying threat model that is based on common attacks against networked systems. The adversary is capable of intercepting communication from the sensor to controller (process measurements), and controller to actuator (manipulated variables). The attacks we consider are categorised as DoS, integrity (man-in-the-middle) and replay attacks. We will investigate the impact of these attacks in terms of measuring the impact on safety and operating cost of the plant. In the following section, we briefly discuss what this means, and explain how the attacks are modelled. 

Past studies have investigated the safety and economic impact of DoS and man-in-the-middle attacks on the TE model \cite{HUANG200973, 10.1145/1966913.1966959, Krotofil2013}. We are following their attack model; however, our focus is to generate optimised combinatorial attacks. We extend their analysis by undertaking a more comprehensive search and examining the possibility of forming combinations of attacks. The attack parameters are as follows: 

\textbf{Attack Targets}: The control strategy selected for our investigation, Larsson et al. in \cite{ie000586y}, uses 16 XMEAS variables, and 9 XMV variables. An adversary may attempt to manipulate signals that are sent from XMEASs to controllers (process measurements), and/or from controller to the XMVs (manipulated variables). The process run time used in this study is 72 hours. An attack begins at time $t_s$ and ends at $t_e$, it can start any time, between start of the plant and the end, $t \in [0-72]$. The manipulated control (XMV) signal $y_{i}^{a}(t)$ and process measurement (XMEAS) $u_{i}^{a}(t)$ are as follows: 

	\begin{equation}
	\begin{array}{l}
    y_i^{(t)}, {for \ t} \notin I_{a} \\
    \hat{y}_i^{(t)}, {for \ t} \in I_{a}
    \end{array}
\end{equation}
			
	\begin{equation}	
  u_{i}^{a}(t)=\begin{cases}
    u_i^{(t)}, {for \ t} \notin I_{a} \\
    \hat{u}_i^{(t)}, {for \ t} \in I_{a}
  \end{cases}
\end{equation}

where $y$ are the $y_{i}^{a}(t)$ and $\hat{u}_i^{(t)}$ are the modified values the adversary sends. 

To investigate the impact of attacks, we considered three types of attacks that could be used: \textit{DoS}, \textit{integrity} and \textit{replay} attacks. 

A \textbf{DoS attack}, is an interruption attack in which a signal is not received by its intended destination. For example, let $y_i$ be the output of sensor $i$ at time $t$, and let $u_i$ be the output from the controller to actuator signal $i$ at time $t$.  When the attack occurs, the response strategy for the controller or the actuator is to use the last received value as the current reading: 

\begin{equation}	
\begin{array}{l}
     \hat{y}_{i}^{(t)}= y_i(t_{s-1})   \\
     \hat{u}_{i}^{(t)}=u_i(t_{s-1}) 
\end{array}
\end{equation}
				
where $\hat{y}_{i}^{(t)}$ and $\hat{u}_i^{(t)}$ are the modified values the adversary sends. 

An \textbf{integrity attack} involves an attacker manipulating signals by changing their values. A na\"ive attacker may listen to the transmitted values and modify them so that they are still within the ranges of possible plant values, since this has the potential to cause some damage. One way to achieve this is to try to modify the sensor measurements $(y_i)$ and manipulated variables $(u_i)$ using observed upper minimum ($integrity_{min}$) and lower maximum ($integrity_{max}$) values: 

An $integrity_{min}$ is where the actual output of the sensor or controller signal $i$ at time $t$ is replaced with a minimum value:

\begin{equation}	
\begin{array}{l}
      \hat{y}_{i}^{(t)} = \underset{t \in T}{min}(y_i(t))  \\
     \hat{u}_{i}^{(t)}= \underset{t \in T}{min}(u_i(t)) 
\end{array}
\end{equation}

An $integrity_{max}$ is where the actual output of the sensor or controller signal $i$ at time $t$ is replaced with a maximum value: 

\begin{equation}	
\begin{array}{l}
     \hat{y}_{i}^{(t)} = \underset{t \in T}{max}(y_i(t))  \\
     \hat{u}_{i}^{(t)} = \underset{t \in T}{max}(u_i(t)) 
\end{array}
\end{equation}

A \textbf{replay attack} involves forging sensor measurements or manipulated variables as in the integrity attack but, this time, it repeatedly replays legitimate data it observed earlier: 
\begin{equation}	
\begin{array}{l}
   
     y_{i}^{r} = [y_{i}^{(r_{start})} \ \textrm{,...,} \ y_{i}^{(r_{end})}] \\
     
     u_{i}^{r} = [u_{i}^{(r_{start})} \ \textrm{,...,} \ u_{i}^{(r_{end})}] \\
     
     \hat{y}_{i}^{(t)} = \hat{y}_i^{r}[t \  \textrm{mod} \ \textrm{len} \ y_{i}^{r}]  \\
    \hat{u}_{i}^{(t)}=  \hat{u}_i^{r}[t \ \textrm{mod} \ \textrm{len} \ u_{i}^{r}]  \\
\end{array}
\end{equation}

where $y_{i}^{r}$ and $u_{i}^{r}$ are the signals recorded by the adversary from the replay period, $r_{start}$ to $r_{end}$. 
\section{Methodology}

\subsection{Baselines}
\label{baselines}
To provide a baseline cost for normal operation of TE plant, 1000 independent runs were executed without disturbances. For this, and all subsequent runs, the plant was operated in Mode 1, the most commonly used configuration in the literature \cite{RICKER1995949}. The following plant operating costs were obtained:

\begin{tabular}{l|c|c}
  \centering
Plant operation  & Max (\$) & Mean (\$) \\
\hline
 Normal (without disturbances)  & 8,218 & 8,208 
\end{tabular}

To inform later $integrity_{min}$ and $integrity_{max}$ attacks, the minimum and maximum values of the XMEAS signals and XMV signals observed under normal operating conditions were recorded. 

\subsubsection{Single random attacks}
To provide a baseline against which to compare the impact of evolved attacks, 500 random attacks were launched on each of the XMEAS and XMV signals per attack type. Attacks were started at hour 2 with the intention of running them for the remainder of the simulation time (i.e. 70 hours). Table \ref{tablexmeas} and Table \ref{tablexmv} shows the impact of each attack in terms of the operating cost of the plant and safety (fastest shutdown time). The variable name and range column describes the name of the sensor or actuator, and the observed minimum and maximum values, which are also used to devise the integrity attacks. The \textit{shutdowns} column denotes the total number of times the plant shut down as a consequence of the attack out of 500 runs. The \textit{shutdown range} column indicates the time at which the plant shuts down after the attack was started. Our experiments show that the fastest attack that can bring the plant down are an $integrity_{min}$ attack on A and C feed flow (XMEAS 4), requiring an attack to last for a duration of 0.52-0.65 hours (31.2-39 minutes), and an $integrity_{max}$ attack on reactor temperature (XMEAS 9), resulting in a shut down at between 0.59-0.63 hours (35.4-37.8 minutes). In these cases, the controller receives fake values from XMEAS 4 and XMEAS 9, which are lower for XMEAS 4 and higher for XMEAS 9 than the legitimate values, and calculates the control values that are sent to the actuators using these fake values. This behaviour causes a significant increase in the reactor pressure, and the plant shuts down as a result. 

\begin{table*} 
  \centering
    \resizebox{0.94\textwidth}{!}{%
    \begin{tabular}{lllcccc}
Variable Number & Variable Name and Range & Attack & Max Cost & Mean Cost & Shutdowns & Shutdown Range(hrs) \\
\hline 
XMEAS 1 &  A-Feed (stream 1) & Max  &  8449  &  8407  &  0  &  -  \\
 & 0.25-0.27 kscmh & Min  &  8364  &  8260  &  0  &  -  \\ 
 &&   DoS  &  8447  &  8331  &  0  &  -  \\
 &&    Replay  &  8429  &  8316  &  0  &  -  \\ \hline
XMEAS 2 & D Feed (stream 2) &  Max  &  1158  &  1152  &  500  &  3.43-3.48  \\
  &  3579.20-3744.46 kgh$^{-1}$ &  Min  &  915  &  902  &  500  &  4.31-4.4  \\
  &&  DoS     &  3149  &  1761  &  500  &  6.9-21.86  \\
  &&  Replay  &  3897  &  2765  &  500  &  13.64-30.42  \\ \hline
XMEAS 3 & E Feed (stream 3)  &  Max  &  739  &  730  &  500  &  2.66-2.72  \\
  & 4339.06-4536.27  kgh$^{-1}$ &  Min  &  1320  &  1312  &  500  &  4.17-4.22  \\ 
  &&  DoS     &  2480  &  1494  &  500  &  3.57-18.16  \\
  &&  Replay  &  3350  &  2330  &  500  &  11.65-22.9 \\ \hline
XMEAS 4 & A and C Feed (stream 4)  &  Max & 384  &  378  &  500  &  1.25-1.34  \\
  & 8.98 9.48 9.24 kscmh   &  Min  &  392  &  361  &  500  &  \textbf{0.52-0.65}  \\ 
  &&  DoS     &  1318  &  743  &  500  &  1.38-6.48  \\
  &&  Replay  &  1227  &  825  &  500  &  2.32-5.99  \\ \hline
XMEAS 5 & Recycle flow (stream 8)  & Max  &  2099  &  2040  &  500  &  8.15-8.42  \\
  &  31.32-33.13 kscmch &  Min  &  3456  &  3415  &  500  &  10.58-10.78  \\ 
  &&  DoS     &  21942  &  10675  &  322  &  10.4-69.7  \\
  &&  Replay  &  \textbf{22429}  &  9169  &  3  &  64.67-64.67 \\ \hline  
XMEAS 7 & Reactor pressure  & Max  &  \textbf{24507}  &  24468  &  0  &  - \\
  &  2793.54-2806.12 kPa  &  Min  &  671  &  650  &  500  &  8.37-8.78  \\
  &&  DoS     &  \textbf{24299}   &  10430  &  254  &  9.01-60.3  \\
  &&  Replay  &  23889  &  10894  &  233    &  9.73-66.15  \\ \hline
XMEAS 8 & Reactor level  &  Max  &  381  &  375  &  500  &  2.81-2.87  \\
  & 62.77-67.24 \% &  Min  &  1017  &  1008  &  500  &  2.83-2.89  \\ 
  &&  DoS  &  7435  &  1989  &  494  &  3.77-35.79  \\
  &&  Replay  &  \textbf{11302}  &  5058  &  418  &  15.67-67.34  \\ \hline 
XMEAS 9  & Reactor temperature &  Max  &  364  &  356  &  500  &  \textbf{0.59-0.63} \\
  & 122.85-122.95 \degree C  &  Min  &  377  &  366  &  500  &  1.23-1.28  \\ 
  && DoS  &  10464  &  1554  &  489  &  0.92-61.64  \\
  &&  Replay  &  \textbf{10774}  &  3681  &  467  &  2.35-67.04  \\ \hline
XMEAS 10 & Purge rate (stream 9)  &  Max  &  8228  &  8195  &  0  &  -  \\
   & 0.1545-0.2689 kscmch &  Min  &  8246  &  8218  &  0  &  -  \\ 
   &&  DoS  &  8235  &  8201  &  0  &  -  \\
   &&  Replay  &  8236  &  8201  &  0  &  -  \\ \hline
XMEAS 11 & Product separator temperature & Max  &  \textbf{10427}  &  10364  &  0  &  -  \\
   & 91.46-92.12 \degree C   &  Min  &  10444  &  10358  &  0  &  -  \\
   &&  DoS  &  10386  &  10270  &  0  &  -  \\
   &&  Replay  &  \textbf{10393}  &  10264  &  0  &  -  \\ \hline
XMEAS 12 &  Product separator level  &  Max  &  896  &  888  &  500  &  5.85-5.95  \\
    & 45.28-54.74 mol \% &  Min  &  1256  &  1245  &  500  &  8.57-8.68 \\ 
   &&  DoS  &  8210  &  3816  &  463  &  8.38-66.43  \\
   &&  Replay  &  8217  &  7802  &  143  &  39.66-69.96  \\ \hline
XMEAS 14 & Product separator underflow  &    Max  &  1370  &  1357  &  500  &  9.09-9.79  \\
  & 51.64-55.89  \%  &  Min  &  1449  &  1385  &  500  &  9.71-10.41  \\ 
  &&  DoS  &  4038  &  2274  &  500  &  11.07-32.98  \\
  &&  Replay  &  4431  &  3004  &  500  &  19.17-36.51  \\ \hline 
XMEAS 15 & Stripper level  & Max  &  1756  &  1740  &  500  &  13.48-13.71  \\
  & 45.29-54.57 \% &   Min  &  1804    &  1793  &  500  &  13.31-13.54  \\ 
  &&  DoS  &  8234  &  5368  &  413    &  19.77-69.61  \\
  &&  Replay  &  8234  &  8181  &  16  &  57.03-68.13  \\ \hline
XMEAS 17 & Stripper underflow (stream 11) &   Max  &  312  &  305  &  500  &  \textbf{1.02-1.03} \\
  & 22.37-23.41 m$^{3}$ h$^{-1}$ &  Min  &  387  &  379  &  500  &  1.01-1.02  \\
  &&  DoS     &  4512  &  2298  &  500  &  6.68-37.75  \\
  &&  Replay  &  5716  &  4312  &  500  &  28.45-48.24  \\ \hline
XMEAS 31 & C in Purge  &  Max  &  \textbf{20515}   &  20438  &  500  &  65.65-66.1  \\
  &  11.87-14.22 mol\% &  Min  &  13493  &  13455  &  500    &  50.02-50.4  \\ 
  &&  DoS     &  \textbf{17205}  &  9841  &  0  &  -  \\
  &&  Replay  &  10951  &  8818  &  0  &  -  \\ \hline
XMEAS 40 & G in product  &  Max  &  8312  &  8298  &  0  &  -  \\
 & 51.64-55.89  mol\% &  Min  &  8117  &  8106  &  0  &  -  \\
  &&  DoS  &  8267  &  8208  &  0  &  -  \\
  && Replay  &  8268  &  8212  &  0  &  -  \\ \hline
\end{tabular}
}
\caption{Impact of random attacks on XMEAS variables (500 runs for each attack)}
\label{tablexmeas} 
\end{table*}

\begin{table*} [ht]
  \centering
    \resizebox{1.0\textwidth}{!}{\begin{tabular}{lllcccc}
Variable Number & Variable Name and Range & Attack & Max Cost & Mean Cost & Shutdowns & Shutdown Range(hrs) \\
\hline 
XMV 1 &  D feed flow (stream 2)  &  Max  &  8209  &  8198  &  0  &  -  \\
& 62.89-63.12 kgh$^-1$   & Min  &  8226  &  8217  &  0  &  -  \\ 
  &&  DoS  &  8216  &  8204  &  0  &  -  \\
  && Replay  &  \textbf{9347}  &  8223  &  0  &  -  \\ \hline
XMV 2 &  E Feed  (stream 3)   &  Max  &  8234  &  8221  &  0  &  -  \\
& 52.99-53.24 kgh$^-1$ & Min  &  8204  &  8194  &  0  &  -  \\ 
  &&  DoS  &  8215  &  8204  &  0  &  -  \\
  && Replay  &  8216  &  8203  &  0  &  -  \\ \hline
XMV 3 & A Feed (stream 1) &  Max  &  8352  &  8342  &  0  &  -  \\
 & 25.12-27.045 kscmh  & Min  &  8259  &  8248  &  0  &  -  \\
  &&  DoS  &  8277  &  8216  &  0  &  -  \\
  && Replay  &  8247  &  8210  &  0  &  -  \\ \hline
XMV 4 &  A and C Feed (stream 4)  &  Max  &  \textbf{10101}  &  10085  &  0  &  -  \\
& 59.93-61.32  & Min  &  9339  &  9290  &  500  &  38.49-39.08  \\ 
  &&  DoS  &  \textbf{14972}  &  8433  &  2  &  68.09-68.09  \\
  && Replay  &  8873  &  8248  &  0  &  -  \\ \hline
XMV 6 &  Purge valve (stream 9)  &  Max  &  \textbf{9552}  &  9542  &  0  &  - \\
& 19.39-32.76 \%   & Min  &  7223  &  7215  &  0  &  - \\ 
  &&  DoS  &  \textbf{9064}  &  8221  &  0  &  -  \\
  && Replay  &  8876  &  8234  &  0  &  -  \\ \hline
XMV 7 &  Separator pot liquid flow (stream 10)  &  Max  &  2382  &  2313  &  500  &  17.65-18.92  \\
& 37.21-37.46 m$^3$ h$^-1$  & Min  &  3355  &  3273  &  500  &  25.83-27.35  \\ 
  &&  DoS  &  8217  &  7578  &  131  &  26.01-69.42  \\
  && Replay  &  8217  &  7975  &  82  &  44.05-68.92  \\ \hline
XMV 8 &  Stripper liquid product flow (stream 11) &  Max  &  2014  &  1962  &  500  &  15.18-15.99  \\
& 46.36-46.55 m$^3$ h$^-1$  & Min  &  2097  &  2060  &  500  &  15.38-16.11  \\ 
  &&  DoS  &  8231  &  5698  &  366  &  22.32-68.72  \\
  && Replay  &  8235  &  6387  &  356  &  24.54-69.52  \\ \hline
XMV 10 &  Reactor cooling water flow &  Max  &  786  &  701  &  500  &  1.65-2.01  \\
& 35.46-36.33 m$^3$ h$^-1$\  & Min  &  \textbf{11703}  &  6651  &  489  &  18.88-67.49  \\ 
  && DoS  &  6093  &  1976  &  500  &  6.18-34.73  \\
  && Replay  &  5909  &  2283  &  500  &  6.4-34.61  \\ \hline
XMV 11 & Condenser cooling water flow  &  Max  &  325  &  319  &  500  &  1.59-1.6  \\
& 5.20-19.69 m$^3$ h$^-1$  & Min  &  414  &  408  &  500  &  \textbf{0.64-0.65} \\
  &&  DoS  &  10803  &  2268  &  477  &  1.61-54.62  \\
  && Replay  &  \textbf{11596}  &  7782  &  108  &  38.32-69.49  \\ \hline
    \end{tabular}}
\caption{Impact of random attacks on XMV variables (500 runs for each attack)}
\label{tablexmv} 
\end{table*}

As reported in Table \ref {tablexmeas}, the experiments carried out showed the following single attacks against process measurements (XMEAS) signals increased the operating cost of the plant significantly: i) $integrity_{max}$ attack on reactor pressure (XMEAS 7) increased the operating cost to \$24,507; ii) $integrity_{max}$ attack on the sensor measuring component C in purge (XMEAS 31) increased the operating cost to \$20,515; iii) DoS attack on reactor pressure (XMEAS 7) increased the operating cost to \$24,299; iv) replay attack on recycle flow (XMEAS 5) increased the operating cost to \$22,429; and v) DoS attack on C in purge (XMEAS 31) increased the operating cost to \$17,205. 

Table \ref{tablexmv} shows the impact of attacking the manipulated variables issued by the controller to actuators (XMVs). The attack that caused the fastest damage is the $integrity_{min}$ attack on the condenser cooling water flow (XMV 11), resulting in a shutdown time of 0.64 hours (38.4 minutes) due to low separator liquid level. Carrying out a $integrity_{max}$ attack on reactor cooling water flow (XMV 10) is able to shut down the plant in 1.65 hours (99 minutes) due to high reactor pressure. Integrity attacks on A and C feed flow (XMV 4), purge valve (XMV 6) and reactor cooling water flow valve (XMV 10) have the potential to increase the operating cost; however, the effect (\$9,064-11,596) is smaller than the attacks on XMEAS, as shown in Table \ref{tablexmv}. 

\subsection{Evolutionary multiobjective optimisation approach}
Due to the computationally intensive nature of this work, and the slowness of the selected TE model simulator in MATLAB, we selected only two of the well-known Pareto-based EMO algorithms, the Non-Dominated Sorting Genetic Algorithm II (NSGA-II) \cite{Deb2000, 996017} and Strength Pareto Evolutionary Algorithm 2 (SPEA2) \cite{Zitzler01spea2:improving}. Both algorithms were implemented using the Distributed Evolutionary Algorithms in Python (DEAP) \cite{Fortin:2012:DEA:2503308.2503311} library. Using the selected algorithms we investigate the following three optimisation problems: 
 \begin{enumerate}
 \item \textbf{Shutdown attacks} is defined as a two objective optimisation problem: minimise time required to shutdown the plant (f1) and minimise the effort required to carry out the attack (f2). Time required to shutdown the plant is the length of the plant operating after attack started. Plant shuts down as a result of exceeding plant operating constraints. We define effort as the total number of sensors and actuators being attacked. 

 \item \textbf{Operating cost attacks} defined as a two objective optimisation problem: maximise the total operating cost of the plant (f1), and minimise the effort required to bring the plant down (f2). 

 \item \textbf{Attacks against detection} defined as two/three objective optimisation: maximise economic loss (f1), minimise detection (alarm) probability (f2), and minimise detection probability (f3). 
 \end{enumerate}

\subsubsection{Detection methods}
The TE model generates a total of 51 data variables, 41 process measurements (XMEAS) and 12 manipulated variables (XMV). This data was used to train the detection methods. To detect attacks, three supervised learning methods -- decision tree (CART, tree depth=50), AdaBoost (with CART, number of estimators=100), random forest (with CART, number of estimators=25) -- and one unsupervised learning method -- one-class SVM (kernel=RBF, $\upsilon=0.00346$ and $\gamma=0.018$) -- were used. The training data for supervised learning were selected from $integrity_{min}$ and \textit{$integrity_{max}$} attacks on measured variables (XMEAS). Attack samples were generated by carrying out integrity attacks on XMEAS signals with durations of between 20 minutes and 3 hours. The dataset used for training the unsupervised learning method consists of normal operational data without any attacks. The performance of the detection methods were measured using F1 score and false positive rate:

\begin{equation}
F_1  = 2{\frac{precision \times {recall}}{precision + recall}}
\end{equation}
\begin{equation}
\text{False Positive Rate (FPR)}  = \frac{FP}{FP + TN}
\end{equation}

where FP is false positives and TN is false negatives. 

The performance of the detection algorithms on test data, unseen cases of integrity attacks on both XMEAS and XMV are illustrated in Table \ref{tabledetection}.

\begin{figure}[h!]
\centering
\includegraphics[width=0.75\textwidth]{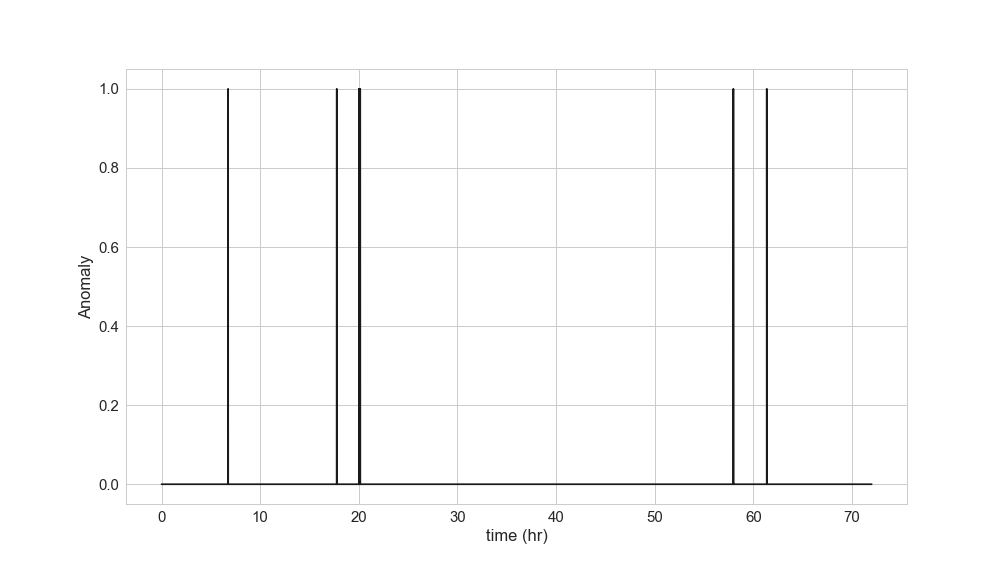}
\caption{Anomaly detection using random forest (under normal operating condition)}
\label{anomaly}
\end{figure}

Each execution of the plant produces a data matrix of size 51 x 36000 data-points (500 points per hour for 72 hours). Systems like the TE plant are prone to natural noise due to behaviour of the physical components of the systems (e.g. actuator and sensors degrading over time, components of the plant wearing, or other kinds of natural noise in the environment). Detection should be robust against natural noise, and distinguish between the normal plant disturbances and attack conditions. Figure \ref{anomaly} shows random forest classifying data-points for a normal execution of the plant, under no attack, for 72 hours. False positives are the lines pointing at 1. 

\begin{table} [!h]
\centering
  \begin{tabular}{ l | l l }
    \hline
     Algorithm & F1 & FPR \\ \hline	
     Decision Tree &  87.74 & 0.19  \\
     Random Forest &  90.54 & 0.15 \\
     AdaBoost  & 72.40 & 0.016 \\
     One-Class SVM & 86.62 &  0.40 \\
  \end{tabular}
\caption{Performance of detection mechanisms}
\label{tabledetection} 
\end{table}

Declaring that an attack is taking place at present requires the detection to be robust to false positives. To cater for this, we use a sliding window of size 100 to declare that an attack is present only if the percentage of anomalous data points in a window exceeds a threshold to ensure false positives do not overwhelm the operator. 
\begin{figure}[h]
\centering
\includegraphics[width=0.75\textwidth] {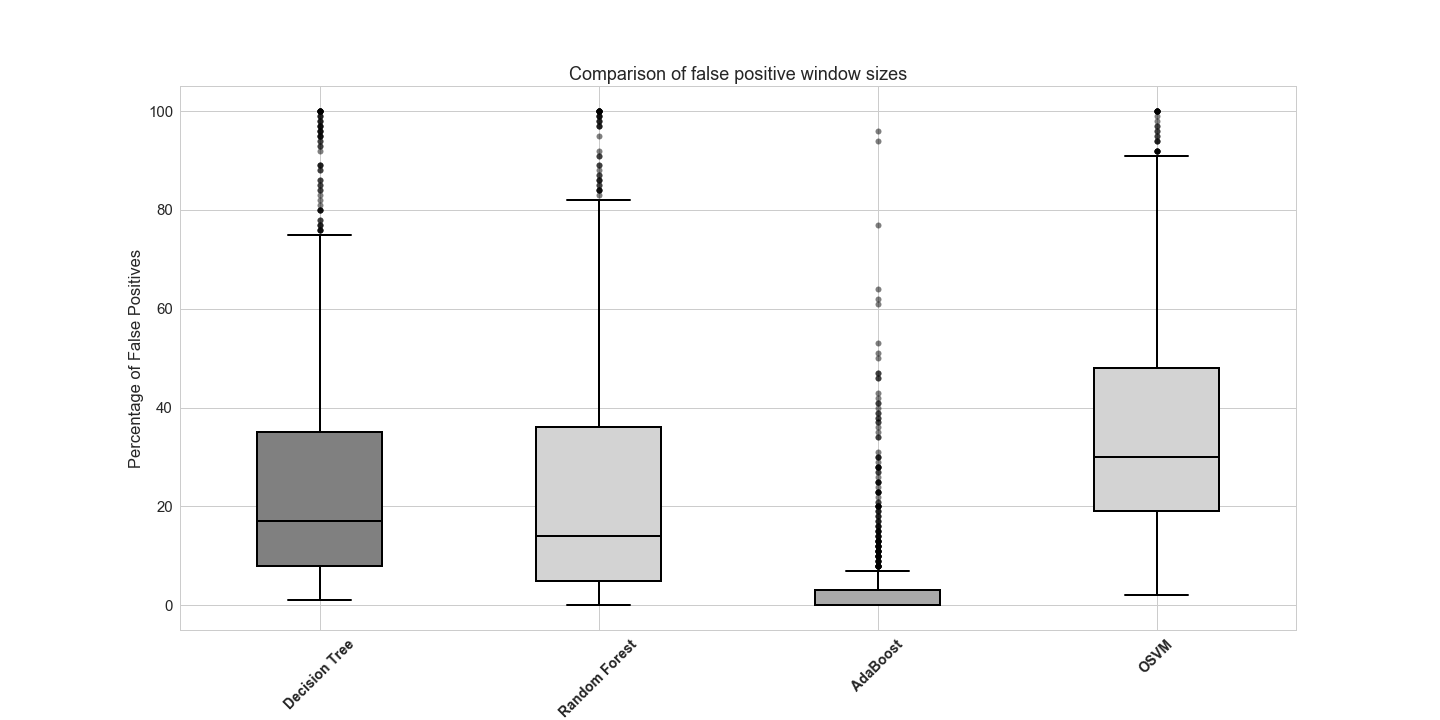}
\caption{Percentage of false positives for detection methods in a window size of 100 (under normal operating conditions)}
\label{windowplot}
\end{figure}

We executed the TE model 1000 times under normal conditions, without any attacks, using a different random seed for each replicate to ensure that randomness was achieved. Then, using each detection method, we investigated the highest false positive percentage seen for each run. Results obtained are shown in Figure \ref{windowplot}. Based on these results, plant operator will need to define a threshold for raising an alarm or declaring an attack is taking place. For this study, we select the 99\% percentile as the threshold for false alarm: 99\% for decision tree; 99\% random forest; 98\% One-Class SVM; and 47\% for AdaBoost. The objective of the attacker is not to raise any alarms while causing some damage by keeping the probability of attack as low as possible by reducing the number of attack data points in the sliding window below the defined threshold to ensure no alarms are raised.

\subsubsection{Representation}
Individuals (chromosomes) are represented as a list, consisting of 25 positions, each position (gene) denoting a sensor or an actuator. The types of genes are represented as integers, denoting the attack and its starting hour. Initial population of individuals are are generated randomly. 

\subsubsection{Fitness function}
The fitness of individuals is evaluated by converting individuals into MATLAB scripts that is executed on the TE plant. The performance of the shutdown attacks is evaluated based on plant run time (f1) and number of variables attacked (f2). The performance of the economic loss attack is evaluated based on the operating cost (f1) of the plant and number of variables attacked (f2). The performance of the attacks against detection is evaluated based on the  detection probability (f1), increased operating cost of the plant (f2), and number of variables attacked (f3, for 3-objective optimisation). Data collected from TE model is tested against the detection model to determine the detection probability.
 
\begin{table} [h]
\centering
  \begin{tabular}{ |l | l| }
    \hline
    Parameters & Value  \\ \hline	
    Chromosome size & 25 \\
    Representation of genes & Integers  \\
    Number of generations & 500-1000 \\ 
    Parent population ($\mu$)  & 100-400 \\
    Crossover & Two-point crossover \\
    Mutation & Uniform mutation \\
    Selection & NSGA-II, SPEA2 \\
    Crossover probability ($cxpb$) & 0.8-0.90 \\
    Mutation probability ($mutpb$) & 0.05-0.15 \\
    Probability of mutating a gene & 0.05-0.08 \\
\hline	
\end{tabular}
\caption{Operators and parameters for evolutionary multiobjective optimisation}
\label{geneticparamdetection} 
\end{table}

\subsubsection{Evolution} 
Table \ref{geneticparamdetection} shows the genetic parameters and the operators used in our experiments. Once the fitness of the initial population$_{0}$ has been evaluated, the evolutionary loop begins to generate the next generation (gen=1) population, as illustrated in Algorithm \ref{algo_mulambda}.

 \begin{figure}[h]
  \centering
  \begin{minipage}{.7\linewidth}
\begin{algorithm}[H]
   \newcommand{\forcond}{$i=0$ \KwTo $\mu$}
  \SetKwFunction{FNSGA}{NSGA-II}
  \SetKwFunction{FSPEA}{SPEA2}
    \SetKwFunction{FvarOr}{vary}
  \SetKwProg{Fn}{Function}{:}{}
  \Fn{\FNSGA{$\mu$, $mutp$, $cxpb$}}{
       $ParetoFront$=[]\;
         $\mu$=pop size,
        $pop$=generateRandomPop($\mu$)\;
        $pop$=evaluateFitness($pop$)\;
        $gen$=1\;
       \While{$gen\leq ngens$}{
        $offspring$=selTournament($pop$,$\mu$)\;
        $offspring$=crossover($offspring$,$cxpb$)\;
        $offspring$=mutate($offspring$,$mutp$) \;
        $offspring$=evaluateFitness($offspring$)\;
	    $pop$=selectNSGA2($offspring$+$pop$,$\mu$)\;
        $ParetoFront$.update($pop$)\;
        $gen$=$gen+1$\;
    } 
\emph{return ParetoFront}\;   }
  \SetKwProg{Fn}{Function}{:}{}
  \Fn{\FSPEA{$\mu$, $mutp$, $cxpb$}}{
     $ParetoFront$=[]\;
     $\mu$=pop size,
     $pop$=generateRandomPop($\mu$)\;
     $pop$=evaluateFitness($pop$)\;
     $gen$=1\;
       \While{$gen\leq ngens$}{
        $offspring$=vary($pop$,$\mu$,$mutpb$,$cxpb$)\;
        $offspring$=evaluateFitness($offspring$)\;
	    $pop$=selectSPEA2($offspring$+$pop$,$\mu$)\; 
	    $ParetoFront$.update($pop$)\;
        $gen$=$gen+1$\;
    } 
\emph{return ParetoFront}\;    
 }
   \SetKwProg{Fn}{Function}{:}{}
  \Fn{\FvarOr{$pop$, $\mu$, $mutp$, $cxpb$}}{
        $offspring$=[]\;
       \For{\forcond}{
         $random$=randomGenerator(0,1) \;
         \uIf{random $<$ cxpb}{
            $ind1$,$ind2$=selectTwoParents($pop$)\;
            $child1$,$child2$=crossOver($ind1$,$ind2$)\;
            $offspring$.add($child1$) \;}
  \uElseIf{random $<$ cxpb+mutp}{
      $ind$=selectOneParent($pop$) \;
      $child$=mutate($ind$)\;
      $offspring$.add($child$) \;}
  \Else{
      $ind$ = selectOneParent($pop$) \;
      $offspring$.add($ind$) \;}
    }

 \emph{return $offspring $}\;      
  }
\caption{Evolutionary multiobjective optimisation algorithm for generating attacks}
\label{algo_mulambda}
\end{algorithm}
  \end{minipage}
\end{figure}

 First, the individuals in population$_{0}$ are subject to the genetic variation operators to generate the next population of offspring. For NSGA-II we used the same genetic variation operators as used in the original algorithm \cite{996017}: tournament selection, two-point crossover, and uniform mutation. Crossover and mutation rates were manually tuned based on values that are usually chosen within the literature: cross-over probabilities used are between 0.8-0.9 and mutation probability between 0.05-0.15. For SPEA2, we used the \textit{vary} function shown in Algorithm \ref{algo_mulambda}, where, on each of the $\mu$ iterations, randomly picked individuals are subject to one of the three operations: two-point crossover, uniform mutation, or reproduction. Our initial experiments showed these operators performed better than the evolutionary operators that were used in standard SPEA2 \cite{Zitzler01spea2:improving}: binary tournament selection, single-point crossover, and bit-flip mutation. Functions $selectNSGA2$ and $selectSPEA2$ are the selection operators of  NSGA-II and SPEA2, readily available in DEAP library. Both selection operators use the (parent+offspring) population to select the next generation: in other words the next generation of the population is produced from both the generated offspring and current parent population$_{0}$. 

NSGA-II creates a Pareto rank of individuals from (parent+offspring) population using non-dominated sorting to select the next generation of individuals. If individuals have the same ranking score, then the crowding distance assignment method based on density estimation is used to select the individuals that are in the least crowded regions within the rank. SPEA2 calculates the strength of the individuals based on domination and density information, to select the new population for the next generation. 

The obtained Pareto front set is updated with the new population, and we use the elements in this set to calculate the hypervolume at each generation of the evolution to monitor the convergence speed of the EMO algorithms. All experiments were carried out using DEAP library.

 \subsubsection{Performance metrics for evolutionary multiobjective optimisation}
Multiobjective evolutionary algorithm have three important goals \cite{Zitzler:2000:CME:1108872.1108876}: i) to minimise the distance between the obtained non-dominated solution set and the true Pareto-optimal set ii) to obtain a good, in most cases uniform distribution of solutions; iii) to maximise the extent of the obtained non-dominated solutions. Therefore, a wide variety of performance metrics \cite{7360024} have been proposed to measure and compare the performance of EMO algorithms. In this study, we selected three of most frequently used EMO performance metrics, namely, hypervolume, spread, and inverted generational distance (IGD) with which to compare the performance of EMO algorithms. 

The \textbf{hypervolume} \cite{10.1007/BFb0056872}, also known as the \textit{S-metric} or \textit{Lebesgue measure} is used for comparing convergence and diversity of a Pareto front. It measures the size of volume of the region between the estimated Pareto-optimal front, $P$ and, a reference point $r$. We followed common practice by calculating $r$ as a point defined by taking the worst known values for each of the objectives and shifted it slightly towards some unattainable values to ensure it is placed in a way that will be dominated by everything else. The hypervolume is defined as follows:

\begin{equation}	
\mathrm{{HV(P,r)}}=\mathcal{L}\left( \bigcup ^{|P|}_{i=1} [\mathrm {r}, \mathrm {i}]\right) 
\end{equation}

It is the union of Lebesgue measure $L$ for all points in $P$ with respect to the reference point $r$. A Pareto front with a higher hypervolume is considered to indicate a better performing EMO algorithm. The \textbf{spread} metric $\triangle$ \cite{996017}, also known as the diversity metric, measures the diversity of the solution by calculating Euclidean distance, $\mathrm{d_i}$, between consecutive solutions. $\mathrm{d_f}$ and $\mathrm{d_l}$ are the Euclidean distances between the \textit{extreme} solutions of the true Pareto front and the boundary solution of the obtained Pareto front \cite{996017}. Assuming there are N solutions in the obtained Pareto front, i=1,i=2,..,N-1, $\mathrm{\bar{d}}$ is the average of all distances $\mathrm{d_i}$. Smaller value of $\triangle$ is desired, indicating better spread of solutions. 

\begin{equation}	
{\Delta} =\frac{d_{f}+d_{l} +{\sum}_{i = 1}^{N-1} \left| d_{i}-\bar{d} \right|}{d_{f}+d_{l}+(N-1) \bar{d}}
\end{equation}

    The \textbf{inverted generational distance (IGD)} \cite{870296} is another widely used metric to measure both convergence and diversity. IGD is defined as: 

\begin{equation}
IGD = \frac {\sqrt{ \sum_{i = 1}^{P^*} d(i, P)}} {\left| {P^*}\right|}
\end{equation}

where $\mathrm{P^*}$ denote  the number of solutions in the optimal Pareto front, and $\mathrm{d_i}$ is the Euclidean distance between solution $\mathrm{i}$ and the nearest member in the obtained Pareto front, $\mathrm{P}$ \cite{870296}. A value, near $\mathrm{IGD=0}$ indicates better coverage of Pareto front and near true Pareto front. Due to size of the problem it was not possible to calculate the true Pareto front, and instead, a reference true Pareto front was computed for each problem by aggregating the obtained non-dominated solutions from all runs to obtain a single front. These values were used to estimate the extreme values for the spread metric, and the reference Pareto front for the IGD metric. 

To compare the performance of the EMO algorithms, the Kruskal-Wallis test was used for data that do not fit a normal distribution, and a one-way ANOVA test was used for data with a normal distribution. The confidence interval for all experiments is 95\%.

\subsection{Random generation of combined attacks}
\label{randomcombin}

To determine the effectiveness of using EMO algorithms for attack generation some initial simple experiments were carried using a random generator and compared with EMO approach. As explained previously, there are in total 25 sensors and actuators to attack, and four possible attacks ($DoS$, $integrity_{min}$, $integrity_{max}$ and $replay$). For the comparison study, we decided to keep the search simple by investigating only shutdown attacks. 10 sets of 50,000 randomly generated attack strategies were generated, limiting number of attacks in each strategy to maximum of 7, out of 25. For each set, a new seed was used for the random number generator of the TE model to ensure randomness in the simulation of the plant. Attack type and targets (sensors and actuators to attack) were randomly selected. Attacks were started at hour 2, and they were left to run until the end of the simulation time, (i.e. remaining 70 hours). 

To compare the performance of the EMO against the randomly generated combinatorial attacks, 10 sets of experiments were carried out using SPEA2 algorithm and the genetic operators in Table \ref{geneticparamdetection} with a cross-over probability of 0.9, mutation probability of 0.05, and the independent probability of mutating a gene was 0.05. All experiments started from a random initial population of 100 individuals, and EMO was run for 500 generations. The same seeds used for the random number generator of the TE plant in randomly generated attacks were used here to prevent variability.

\section{Experiments and results}

\subsection{Comparison of random generation and EMO approach}
\label{randomemoa}
In this section, we report the results obtained from experiments to compare the random generation of shutdown attacks with EMO approach. 

\begin{figure} [ht!]
\centering
  \subcaptionbox{Distribution of shutdown attacks generated using random generation}{%
    \includegraphics[width=.47\linewidth]{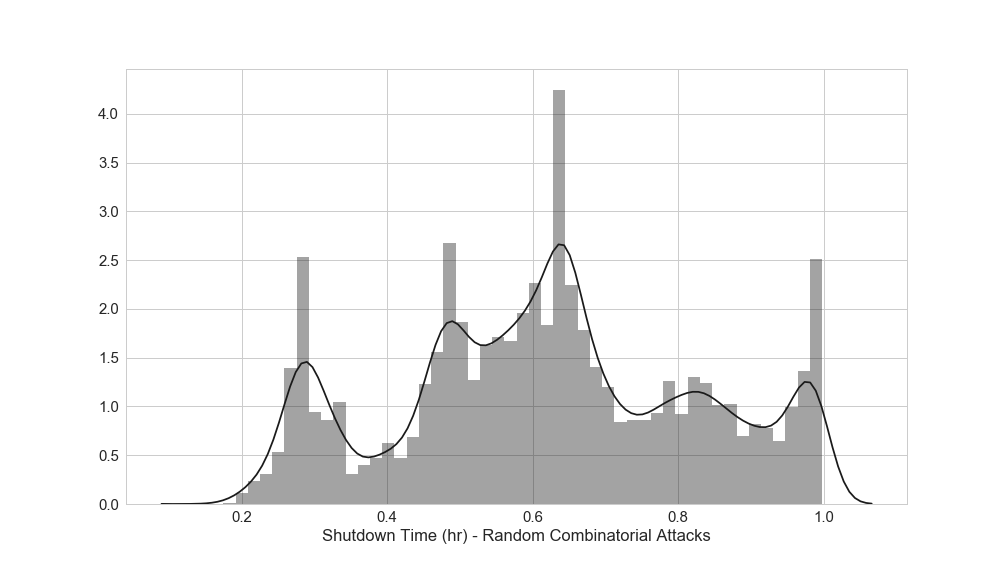}}\quad
     \subcaptionbox{Distribution of shutdown attacks generated using EMO approach}{%
    \includegraphics[width=.47\linewidth]{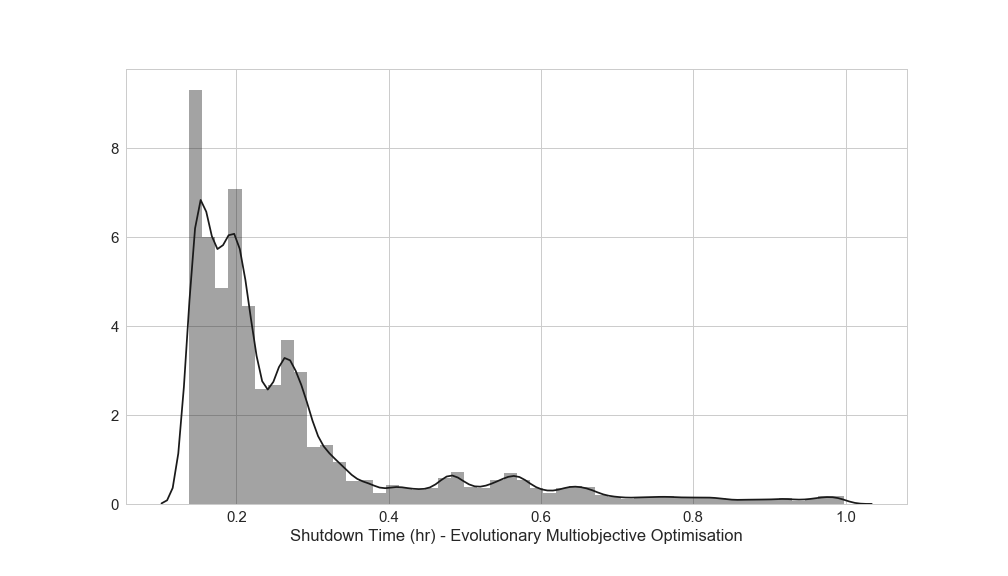}}\quad
  \caption{Comparison of random and EMO approaches}
 \label{randomcombinatorial}
\end{figure}

After removing duplicates, random generation produced a total of 442,125 unique attacks. Just over 18.5\% of these attacks were able to bring the plant down in less than 1 hour. Figure \ref{randomcombinatorial}a illustrates the distribution of these attacks. The best attack strategy random search generated was the attack that shut down the plant in 0.158 hours by attacking 6 sensors and actuators (XMEAS4, XMEAS8, XMEAS10, XMEAS11, XMEAS17, XMV1) using $integrity_{max}$ attack.

 Figure \ref{randomcombinatorial}b show the results obtained using EMO approach. In total, EMO generated 35,658 unique attacks, and 86\% of these attacks were under 1 hour, as indicated by the distribution skewed towards right in Figure \ref{randomcombinatorial}b. These attacks performed far better than those generated by random generation, generated a high number of attacks with shutdown time of 0.138-0.156 hours. These results show the EMO approach is more effective at generating attacks than random chance, and justified our work to use the approach to generate further attacks. The results obtained for generating attacks using both EMO algorithms is presented in the next subsections. 

\subsection{Attacking the safety of the plant: shutdown attacks}
In this section we report the performance of EMO algorithms that targetted the safety of the plant by searching for attacks that shut down the plant by attacking the least number of sensors and actuators. To compare the performance of NSGA-II and SPEA2 algorithm, results were collected over thirty runs for each EMO algorithm using a cross-over probability of 0.9, a mutation probability of 0.05, and the independent probability of mutating a gene was 0.05. For each of the thirty runs of evolution, a new seed was used to produce a different initial random population of size 100. The same sets of seeds were used for NSGA-II and SPEA2 to ensure both algorithms started with the same initial population. Similarly, the seed used for the TE Plant was kept the same for both algorithm. Each evolution was run for 500 generations. All experiments were carried out on a HPC platform facility at the University College London. 

\begin{figure*}
  \centering
  \subcaptionbox{Pareto front for shutdown attacks $<1$ hour (NSGA-II)}{%
    \includegraphics[width=.45\linewidth]{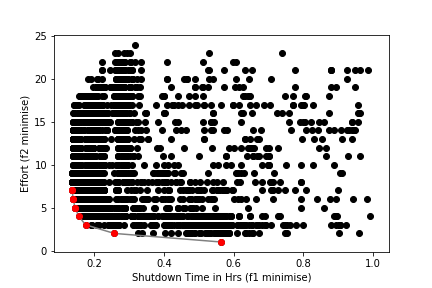}}\quad
    \subcaptionbox{Pareto front for shutdown attacks $<1$ hour (SPEA2)}{%
    \includegraphics[width=.45\linewidth]{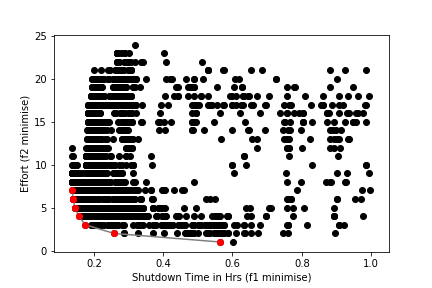}}\quad
     \subcaptionbox{Hypervolume for shutdown attacks}{%
    \includegraphics[width=.45\linewidth]{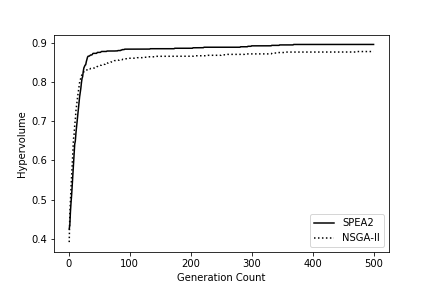}}\quad
       \subcaptionbox{Measurement metrics for shutdown attacks }{%
    \includegraphics[width=.45\linewidth]{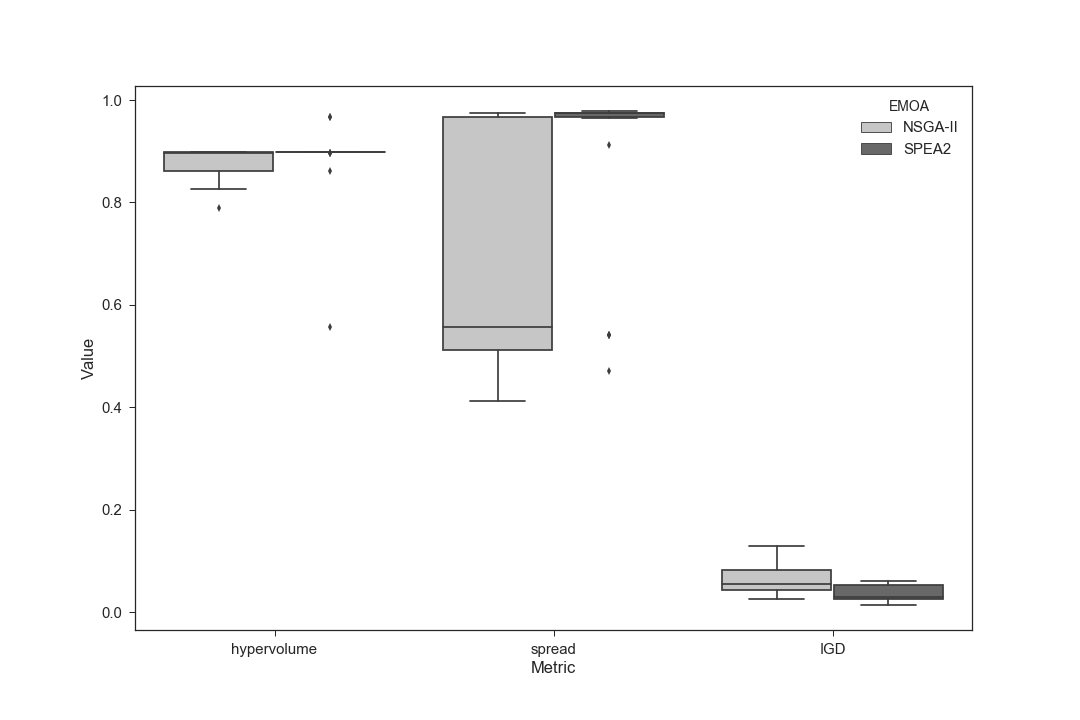}}\quad
    
  \subcaptionbox{Pareto front for operating cost attacks (NSGA-II)}{%
    \includegraphics[width=.45\linewidth]{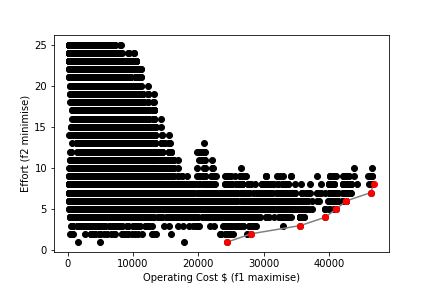}}\quad
    \subcaptionbox{Pareto front for operating cost (SPEA2)}{%
    \includegraphics[width=.45\linewidth]{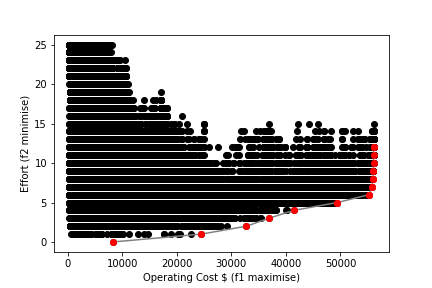}}\quad
     \subcaptionbox{Hypervolume for operating cost attacks}{%
      \includegraphics[width=.45\linewidth]{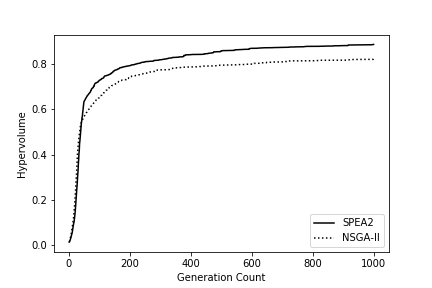}}\quad
          \subcaptionbox{Measurement metrics for operating cost attacks }{%
    \includegraphics[width=.45\linewidth]{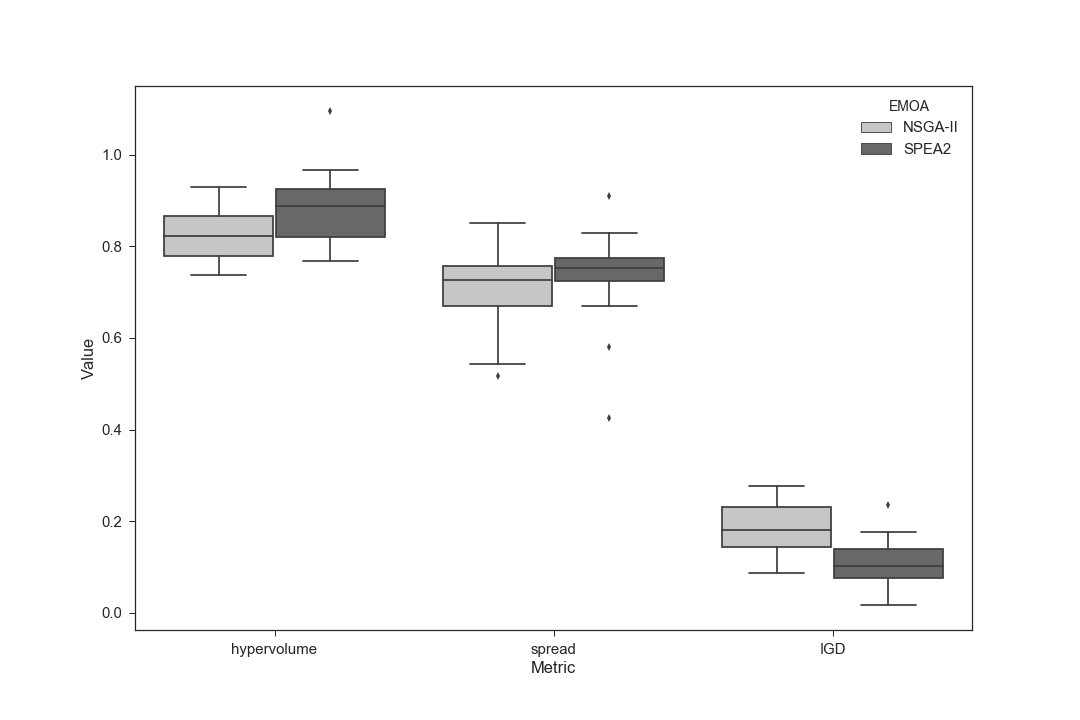}}\quad
  \caption{Results of shutdown and operating cost attacks}
  \label{shutdown}
\end{figure*}
Table \ref{tableops} shows the average and standard deviation of hypervolume, spread and IGD metrics. SPEA2 achieved better hypervolume and IGD with statistical confidence, yielding a better Pareto front and converged faster than NSGA-II. Meanwhile, NSGA-II obtains lower hypervolume and IGD than SPEA2 despite producing significantly better spread. The solutions in the obtained Pareto front were not widely spread across the front, only a small number of signals are required to cause the plant to shut-down. SPEA2 was able search this area better, and converged to a better Pareto front faster than NSGA-II. 

\begin{table}[htbp]
\centering
\begin{tabular}{|l|l|l|l|}
    \hline
     Performance Measure & NSGA-II & SPEA2  & p-value \\ 	\hline
     \textbf{Shutdown Attacks} & & & \\ 
    Hypervolume  & 0.8782 (0.0316) & \textbf{0.8963} (0.0064)  & \textbf{0.000486} \\
          
    Spread   & \textbf{0.6878} (0.2346) &  0.8786 (0.1473) & \textbf{0.000054}  \\
    IGD      & 0.0598 (0.025) &\textbf{0.0368} (0.0154) & \textbf{0.000245} \\
    Fastest Shutdown (hrs) & 0.138 & 0.138 & \\
    
    \hline
    \textbf{Opcost Attacks} & & &\\ 
    Hypervolume  & 0.8235 (0.055)  & \textbf{0.8877} (0.074)  & \textbf{0.003 }\\
    Spread   & \textbf{0.7117} (0.0853)  &  0.7356 (0.095)  & 0.234\\
    IGD      & 0.1861 (0.0554) &  \textbf{0.1110} (0.0494)  & \textbf{0.000260}  \\
    Max. Operating Cost (\$) &  46,814 & \textbf{56,090} & \\
        \hline
     \end{tabular}
\caption{Performance of shutdown and operating cost attacks averaged over all runs}
\label{tableops}
\end{table}

 One of the best run obtained from shutdown attacks using NSGA-II and SPEA2 is shown in Figure \ref{shutdown}a and Figure \ref{shutdown}b. The elements of the final Pareto front obtained at the end of the evolution are plotted in red dots, and some of the members of the Pareto set are shown in Table \ref{tab:spea2}. For this particular run, SPEA2 outperformed NSGA-II in that it found several different attacks with equal fitness functions and effort values in the Pareto set, and it had better Pareto front cardinality; in this case, a total of 12 elements, against the 9 found by NSGA-II. In the plot of the Pareto front some of the points on the diagram refer to multiple attacks with equal fitness; for example, there were 3 attacks using 6 effort and shutdown time of 0.140 hours; 2 attacks with 4 effort and shutdown time of 0.1560 hours. 

\begin{table*}[ht]
    \begin{tabular}{ |p{12cm}| c | c |}
    \hline
    Attack Strategy & Shut-down (hrs) & Effort  \\ \hline
XMEAS4\textsubscript{integrity\textsubscript{min}},XMEAS5\textsubscript{integrityMin},XMEAS8\textsubscript{IntegrityMin},XMEAS11\textsubscript{IntegrityMin}, XMEAS17\textsubscript{IntegrityMin},XMEAS31\textsubscript{IntegrityMin}, XMV6\textsubscript{Replay}, 
& 0.138 & 7 \\ \hline 
XMEAS4\textsubscript{IntegrityMin},XMEAS8\textsubscript{IntegrityMin},XMEAS11\textsubscript{IntegrityMin},XMEAS17\textsubscript{IntegrityMin},
XMEAS31\textsubscript{IntegrityMin}, 
XMV\textsubscript{6\textsubscript{Replay}} & 0.140 & 6 \\ \hline 
XMEAS4\textsubscript{IntegrityMin},XMEAS8\textsubscript{IntegrityMin},XMEAS11\textsubscript{IntegrityMin},XMEAS17\textsubscript{IntegrityMin}, XMEAS31\textsubscript{IntegrityMin} & 0.1460 & 5 \\ \hline
XMEAS4\textsubscript{IntegrityMin},XMEAS8\textsubscript{IntegrityMin},XMEAS11\textsubscript{IntegrityMin}, XMEAS17\textsubscript{IntegrityMin} & 0.1560 & 4 \\ \hline 
XMEAS4\textsubscript{IntegrityMin},XMEAS8\textsubscript{IntegrityMin},XMEAS11\textsubscript{IntegrityMin} & 0.1760 & 3 \\ \hline 
XMEAS8\textsubscript{IntegrityMin},XMEAS11\textsubscript{IntegrityMin} & 0.2579 & 2 \\ \hline 
XMEAS4\textsubscript{IntegrityMin} & 0.5640 & 1 \\ \hline 
Do Not Attack   &  72 & 0 \\ \hline
    \end{tabular}
    \caption{Some elements of the Pareto front for shutdown attack}
        \label{tab:spea2}
\end{table*}

\begin{table*}[ht]
    \begin{tabular}{ |p{12cm}| c | c |}
    \hline
    Attack Strategy & Opcost (\$) & Effort  \\ \hline
XMEAS2\textsubscript{DoS(2,70)},XMEAS7\textsubscript{IntegrityMax(2,70)},XMEAS8\textsubscript{DoS(10,20)},XMEAS11\textsubscript{DOS(30,42)},
XMEAS12\textsubscript{Replay(50,12)},XMEAS15\textsubscript{IntegrityMax(50,12)},XMEAS31\textsubscript{IntegrityMin(30,42)},
XMV1\textsubscript{DOS(10,62)},XMV2\textsubscript{IntegrityMin(2,70)},XMV3\textsubscript{DOS(2,70)},XMV4\textsubscript{DOS(50,12)},XMV11\textsubscript{DOS(50,12)} & 56,090 & 12 \\ \hline  
XMEAS2\textsubscript{DoS(2,70)},XMEAS7\textsubscript{IntegrityMax(2,70)}, XMEAS31\textsubscript{IntegrityMin(30,42)},XMV1\textsubscript{DoS(10,62)}, XMV2\textsubscript{IntegrityMin(2,70)},XMV3\textsubscript{DoS(2,70)} & 55,275 & 6 \\ \hline   
XMEAS2\textsubscript{DoS(2,70)},XMEAS7\textsubscript{IntegrityMax(2,70)}, XMEAS31\textsubscript{IntegrityMin(30,42)}, XMV1\textsubscript{DoS(10,62)}, XMV3\textsubscript{IntegrityMin(2,70)}
 & 
49,449 & 5 \\ \hline  
XMEAS2\textsubscript{DoS(2,70)},XMEAS7\textsubscript{IntegrityMax(2,70)},XMV1\textsubscript{DoS(10,62)},XMV3\textsubscript{IntegrityMin(2,70)}
 & 41,430 & 4  \\ \hline  
XMEAS7\textsubscript{IntegrityMax(2,70)},XMEAS31\textsubscript{DoS(10,62)},XMV3\textsubscript{IntegrityMin(2,70)}& 
36,866 & 3 \\ \hline  
XMEAS7\textsubscript{IntegrityMax(2,70)},XMEAS31\textsubscript{IntegrityMin(30,42)}
 & 
32,761 & 2 \\ \hline  
XMEAS7\textsubscript{IntegrityMax(2,70)}  & 24,479 & 1 \\ \hline 
Do Not Attack   &  8,210 & 0 \\ \hline
    \end{tabular}
    \caption{Some elements of the Pareto front for operating cost attacks using SPEA2}
        \label{tab:multispea2}
\end{table*}

Figure \ref{shutdown}c shows the hypervolume for each of the 500 generations, averaged over all 30 runs. At the end of each generation, the hypervolume was computed according to the Pareto front achieved at that generation to compare the speed of the convergence. For convenience, plotted hypervolume results are normalised to the interval [0-1] according to the best hypervolume value possible, estimated based on the maximum measurement obtained. SPEA2 converges faster to a better Pareto front whereas NSGA-II requires more time to reach a slightly worse Pareto set. This is supported by IGD metric, as shown on the boxplot in Figure \ref{shutdown}d SPEA2 scores a better IGD score. 

Results obtained show combined simultaneous attack generated using EMO approach performed better than single and combined attacks generated randomly against the XMEAS and XMV variables. As discussed in Section \ref{baselines}, it took 0.52 hours to bring down the plant, whereas, EMO found a range of attacks that could bring down the plant faster, in 0.138-0.5640 hours. Random combined attacks (Section \ref{randomemoa}) at the very best produced an attack that could bring down the plant in 0.158 hours.    

\subsection{Causing economic loss: Operating cost attacks}
In this section we report the performance of EMO algorithms that targetted the operating cost of the plant to cause economic loss by attacking the least number of sensor and actuator signals.
Due to the slowness of the fitness function, only a small set of attack start times (2, 10, 20, 30, 50 hours) and attack duration (10, 12, 20, 42, 50, 52, 62, 70 hours) were used to generate attacks. As before, attack types were DoS, replay, $integrity_{max}$ and $integrity_{min}$. Individuals were represented as chromosomes encoded as a list of 25 integers (genes), each position denoting a sensor or an actuator. In total each gene could have a value between 0-37 (0 representing not to attack, remaining 36 representing 9 different attacks per attack type with different start time and attack duration). This is a search space of size $37^{25}$. Twenty runs were carried out for each EMO algorithm to analyse the impact of attacks on the operating cost of the plant. As before, attacks were generated using the same genetic operators, shown in Table \ref{geneticparamdetection}, with a cross-over probability of 0.85, mutation probability of 0.10, and independent probability of mutating a gene was 0.08. Each evolution started with a random population of 400 individuals, and ran for 1000 generations.

One of best runs of operating cost attacks using NSGA-II and SPEA2 is shown in Figure \ref{shutdown}e and Figure \ref{shutdown}f. As with shutdown attacks, SPEA2 performed better than NSGA-II both in the quality of obtained Pareto front set, and the time it took to converge. As indicated in Table \ref{tableops}, the average over all the runs showed that SPEA2 has a hypervolume average of 0.8877, against NSGA-II scoring a hypervolume of 0.8235. Figure \ref{shutdown}h shows the boxplots for measurement metrics. SPEA2 produced significantly higher hypervolume and IGD, but no significant difference was found between the two algorithms for spread. Figure \ref{shutdown}g shows the comparison of hypervolume between NSGA-II and SPEA2, averaged over all runs. These results show that SPEA2, as before, is able to search the space faster and produce a better Pareto front with a higher cardinality. SPEA2 (Table \ref{tab:multispea2}) increased the cost of the operating the plant from an average of \$8,208 to \$56,090, whereas NSGA-II increased the operating cost to \$46,814. 

Table \ref{tab:multispea2} shows some of the elements of the obtained Pareto front set for SPEA2. The numbers in the bracket denotes the start time and duration of the attack, for example XMEAS8\textsubscript{DoS(10,20)} means a DoS attack was carried out against XMEAS 8 starting at hour 10, for a duration of 20 hours. Overall, these results show that EMO approach can be used successfully to generate attacks that could cause economic loss, by identifying the components that increase the operating cost of the plant.

\begin{figure*} 
\centering
 \subcaptionbox{Pareto front for AdaBoost}{%
    \includegraphics[width=.45\linewidth]{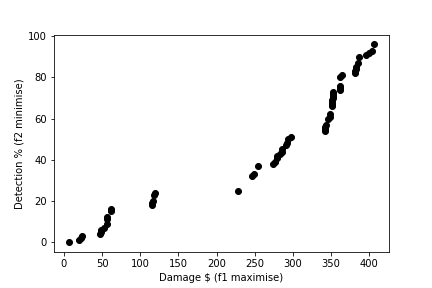}}\quad
 \subcaptionbox{Hypervolume for AdaBoost}{%
    \includegraphics[width=.45\linewidth]{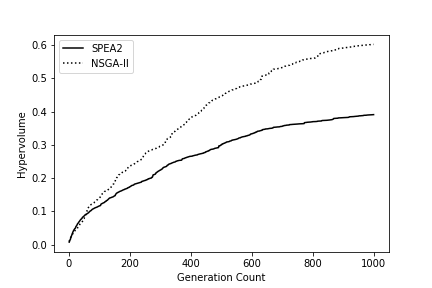}}\quad
    
     \subcaptionbox{Pareto front for decision tree}{%
    \includegraphics[width=.45\linewidth]{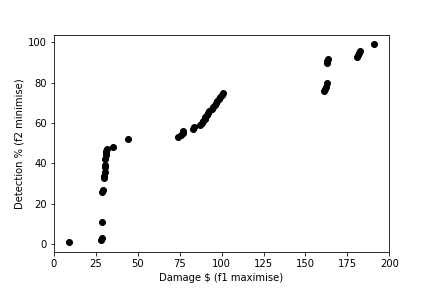}}\quad
 \subcaptionbox{Hypervolume for decision tree }{%
    \includegraphics[width=.45\linewidth]{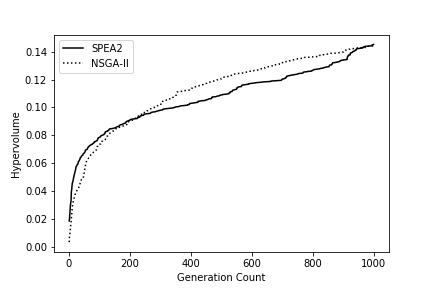}}\quad
    
 \subcaptionbox{Pareto front for random forest}{%
    \includegraphics[width=.45\linewidth]{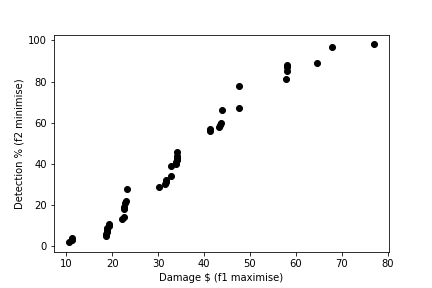}}\quad
 \subcaptionbox{Hypervolume for random forest}{%
    \includegraphics[width=.45\linewidth]{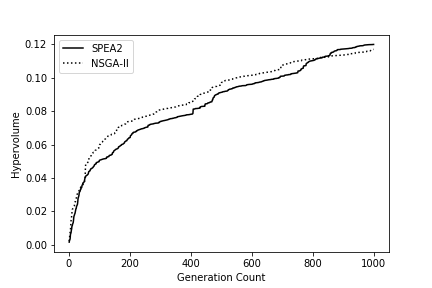}}\quad

  \subcaptionbox{Pareto front for one-class SVM}{%
    \includegraphics[width=.45\linewidth]{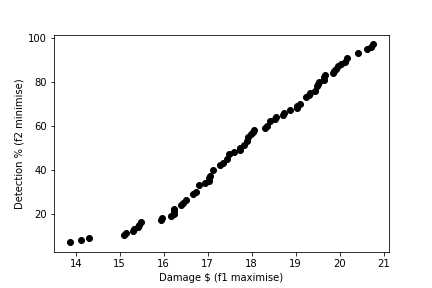}}\quad
      \subcaptionbox{Hypervolume for one-class SVM}{%
    \includegraphics[width=.45\linewidth]{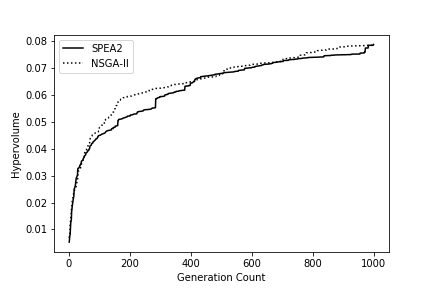}}\quad
    \caption{Pareto front and hypervolume for detection methods, averaged over all runs}
  \label{detection}
\end{figure*}

\begin{figure*} [ht]
 \subcaptionbox{Metrics for AdaBoost}{%
    \includegraphics[width=.5\linewidth]{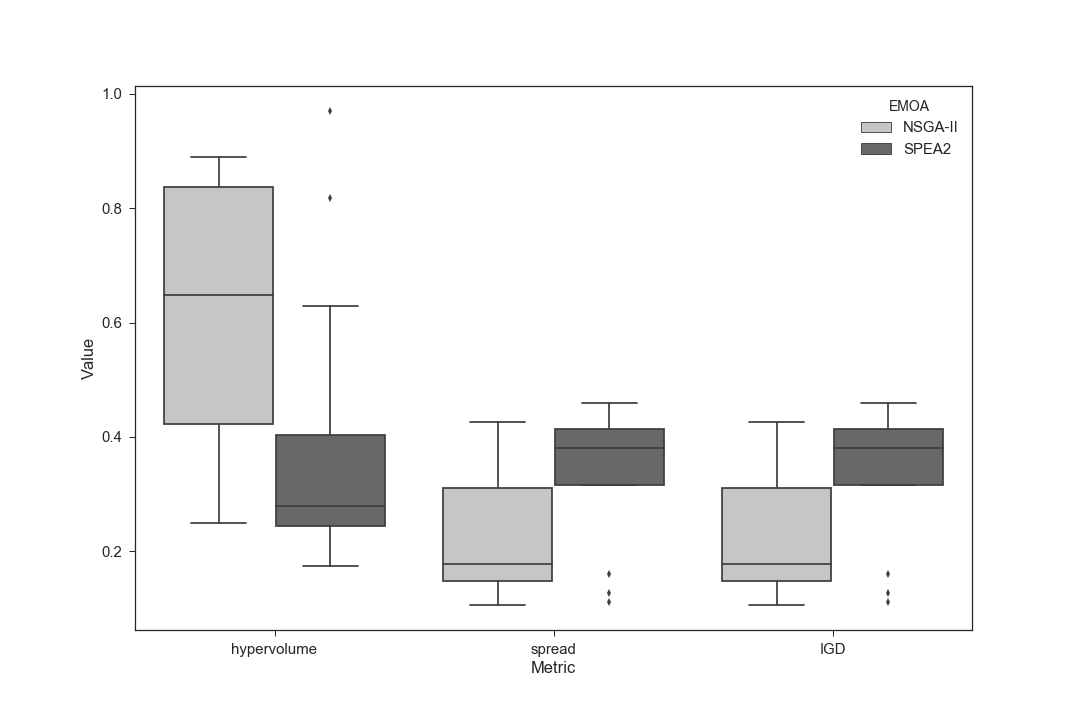}}\quad
      \subcaptionbox{Metrics for decision tree}{
    \includegraphics[width=.5\linewidth]{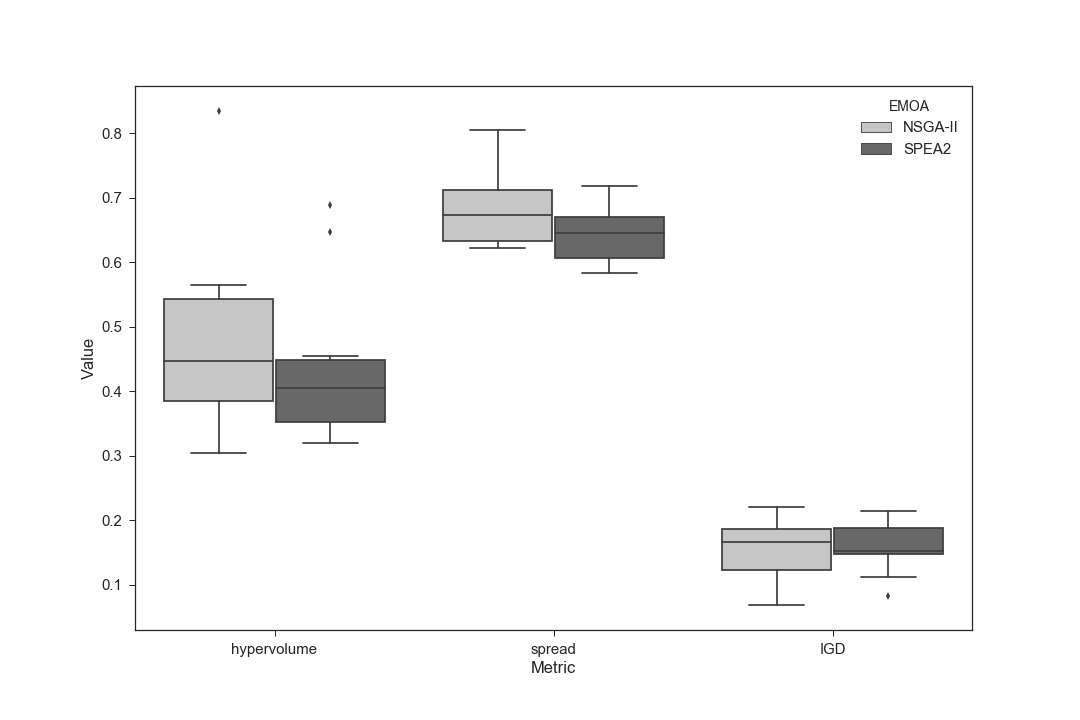}}\quad
  \subcaptionbox{Metrics for random forest}{%
    \includegraphics[width=.5\linewidth]{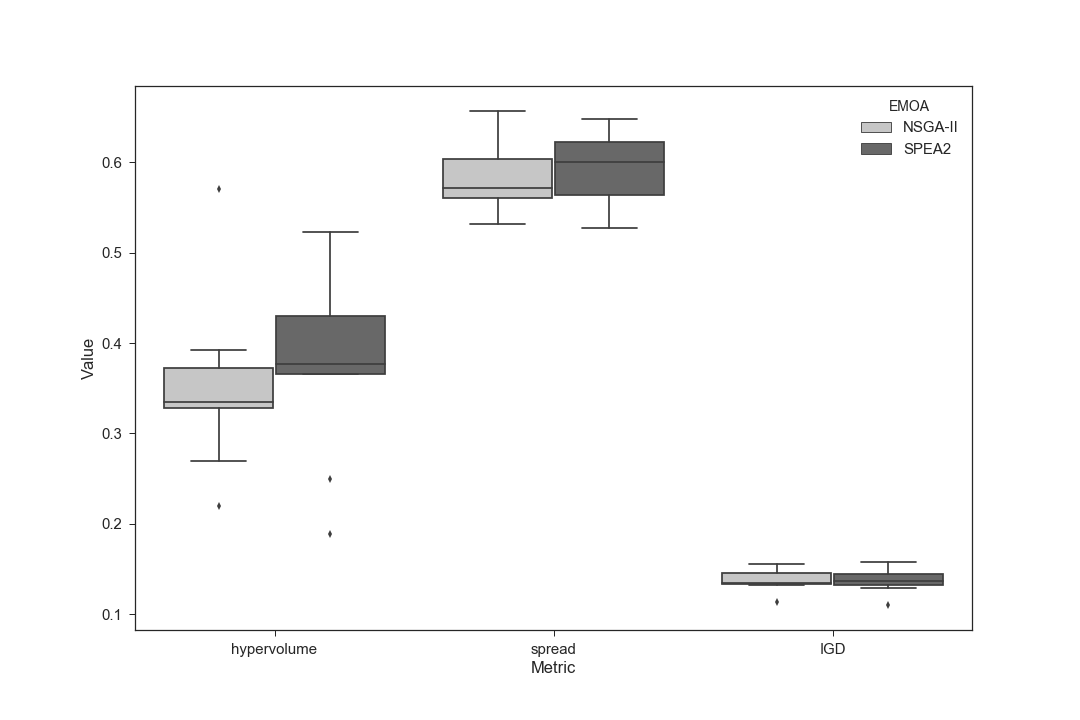}}\quad
    \subcaptionbox{Metrics for one-class SVM}{%
    \includegraphics[width=.5\linewidth]{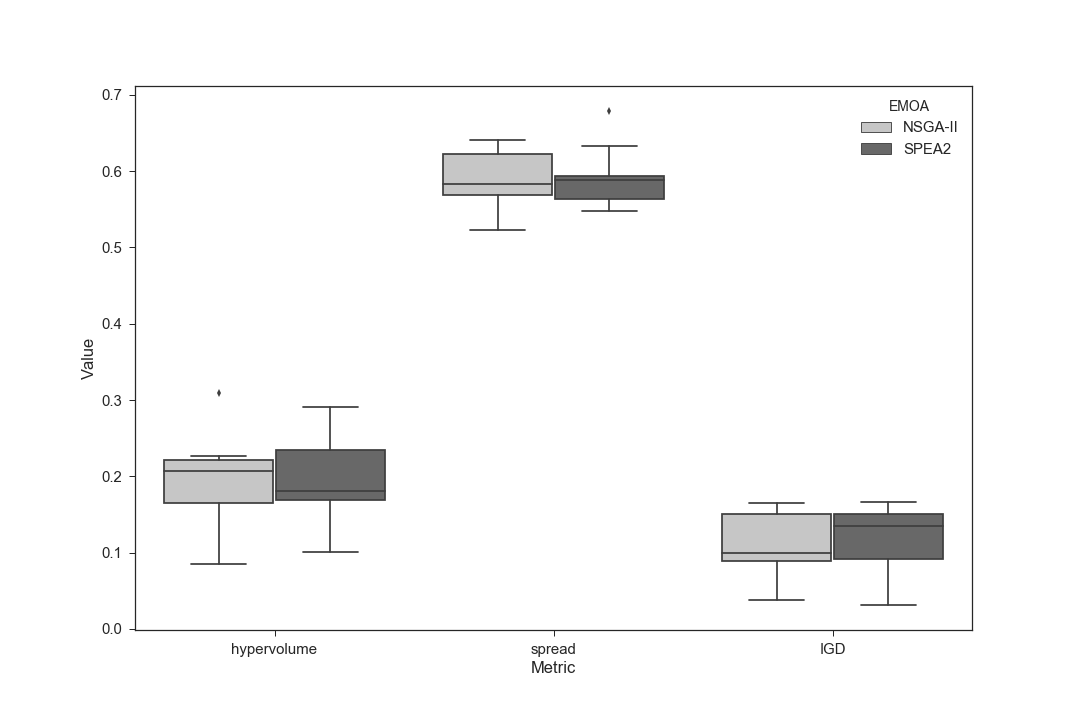}}\quad
\caption{Performance metrics for classifiers, averaged over all runs}
  \label{detection1}
\end{figure*}

\subsection{Generating attacks against detection methods}
As before individuals are represented as a list, consisting of 25 positions, each position denoting a sensor or an actuator. One key obstacle we had with our experiments was time, the execution of the TE model in MATLAB could take up to several minutes. This performance issue influenced the way we encoded our individuals, the size of the population, and the number of generations. To make the problem computationally tractable, we limited the types of genes to a pool of 140 in which half denoted DoS attacks and the remaining half denoted replay attacks. Genes were represented as integers, each number denoting the start time of the attacks (between hour 2-70). The duration of the attacks against detection methods were kept constant for all attacks, 2 hours. This is a combinatorial search problem of size $140^{25}$. Results were collected over 10 runs for each EMO algorithm against each detection method (AdaBoost, decision tree, random forest and one-class SVM) using a cross-over probability of 0.8, mutation probability of 0.15, and the independent probability of mutating a gene was 0.08.

Each evolution started from a random population of 200 individuals, and ran for 1000 generations. Attacks that lead to plant to shutdown were penalised, with a score -1, as our aim was to generate attacks that kept the plant running while causing some damage and evading detection. 

 Despite using a limited number of attack parameters (start time, duration), and using a small number of population and generation size, the obtained result capture some of the weaknesses of the detection methods.

Figure \ref{detection} shows the results obtained. Figure \ref{detection}a, Figure \ref{detection}c, Figure \ref{detection}e and Figure \ref{detection}g show one of the best Pareto fronts obtained against four classifiers. As expected, EMO algorithms were able to exploit low detection rate of AdaBoost, and cause more damage (\$406-6) while keeping the attack detection (alarm) probability at a lower rate. EMO algorithms performed less damaging attacks against decision tree (\$190-10), random forest (\$77-10), and one-class SVM (\$21-13). As before, hypervolume was computed at each generation to analyse the convergence. Figure \ref{detection}b, \ref{detection}d, \ref{detection}f, and \ref{detection}h shows a steady increase of hypervolume suggesting that more generations will yield a better search and convergence. Figure \ref{detection1} shows the hypervolume, spread and IGD boxplots for four classifiers. As indicated in Table \ref{tabledetmetrics} statistical test shows there was no significant difference between NSGA-II and SPEA2 against decision tree, random forest and one-class SVM. However, for AdaBoost, NSGA-II did significantly better, both in terms of obtaining a better Pareto front (as indicated by hypervolume and IGD) metrics and converged a lot of faster as indicated by hypervolume plot (Figure \ref{detection}b). Although, this time SPEA2 had a better spread compared to NSGA-II, it did not yield a better Pareto front.

\begin{table}[t]
\centering
\begin{tabular}{|l|l|l|l|}
    \hline
     Performance Measure & NSGA-II & SPEA2  & p-value \\ 	\hline
     \textbf{AdaBoost} &  & & \\
    Hypervolume & \textbf{0.6020} (0.2257) & 0.3910 (0.2525) &  \textbf{0.032663} \\
    Spread  & 0.7952 (0.0910) & \textbf{0.7165} (0.0841) & \textbf{0.000054}  \\
    IGD    & \textbf{0.2273} (0.1024) & 0.3345 (0.1203) & \textbf{0.0376} \\
    Damage Range (\$) & 15-490.63 & 12.5-424.99 & \\
    \hline
      \textbf{Decision Tree} &  & & \\
    Hypervolume & 0.1442 (0.0464)  & \textbf{0.1452} (0.0416) & 0.8798 \\
    Spread & 0.6825 (0.0564) & \textbf{0.6450} (0.0441) & 0.1340   \\
    IGD    &\textbf{0.1505} (0.0506) & 0.1580 (0.0382) & 0.7284 \\
    Damage Range (\$) & 9-190.63 & 13-182.26 & \\
    \hline
    
    \textbf{Random Forest} & & &\\ 
    Hypervolume & 0.1170 (0.0290) &	\textbf{0.1240 }(0.0304) & 0.2895 \\
    Spread  &\textbf{0.5846 }(0.0384) & 0.5943 (0.0361) & 0.5877 \\
    IGD     & \textbf{0.1375 (0.0110)} & 0.1382 (0.01328) & 0.7622 \\
    Damage Range (\$) &4-76.92 &1.56-56.10 & \\
        \hline
    \textbf{One-Class SVM} & & &\\ 
    Hypervolume & 0.0785 (0.0222) &\textbf{0.0788} (0.0225) & 0.9768\\

    Spread  & \textbf{0.5887} (0.0363) & 0.5909 (0.03789) & 0.8205  \\
    IGD     & \textbf{0.1108} (0.0389) & 0.1181 (0.04144) & 0.7070\\
    Damage Range (\$) & 1.5-21.28 & 2.0-20.75  \\
       \hline
    \end{tabular}
\caption{Performance of detection attacks against classifiers, averaged over all runs}
\label{tabledetmetrics}
\end{table}

As the damage increases, denoted by the increased cost of operating the plant, the probability of detection also increases. EMO algorithms failed to evolve highly damaging attacks against the one-class SVM as it was able to detect DoS and replay attacks better than other detection methods. However, we were able to cause some economic damage with low detection probabilities for other detection methods. 

\subsubsection{Seeding initial population, and generating attacks using 3-objective optimisation}
\label{seeedingsection}

In this section, we report some preliminary experiments that require further investigation, but show promising results. Generating attacks against a strong classifier can be a very time consuming task; this is especially true for cases like our plant model, for which the evaluation of individuals (fitness function) is slow. The slowness of our fitness function also influenced the attack parameters, the size of the population, and the number of generations. To test if we could generate better attacks against decision tree and random forest classifiers, we started some experiments using a seeded initial population that included some good individuals that were obtained previously from experiments carried out against the decision tree classifier (i.e., Figure \ref{detection}c). Seeding is a common practice in single-objective evolutionary algorithms where prior knowledge obtained from previous experiments or expert knowledge is included in the initial population as a good initial estimate,  however advantages and disadvantages of using seeding in EMO algorithms, especially for solving real-world combinatorial optimisation problems require further studies \cite{FRIEDRICH2015223}. 

\begin{figure} [!h]
  \subcaptionbox{Attack generated against decision tree }{%
    \includegraphics[width=.5\linewidth]{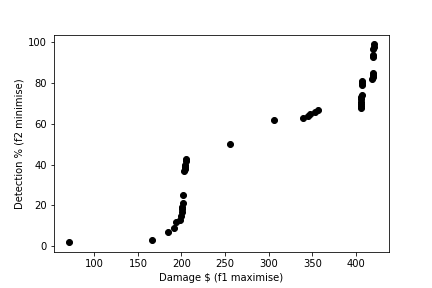}}\quad
     \subcaptionbox{Attack generated against random forest}{%
    \includegraphics[width=.5\linewidth]{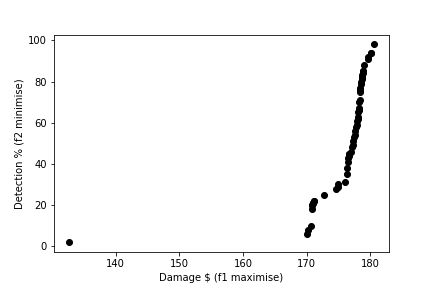}}\quad
  \caption{Seeding population with knowledge gained from prior experiments}
  \label{seeding}
\end{figure}

A comprehensive study is left as future work, and here we report some initial results. We carried out an experiment where we seeded the initial population with 10 of the best individuals obtained from the previous runs of decision tree experiments, and started the EMO from this modified population against the decision tree and random forest classifiers. Figure \ref{seeding}a shows the performance of decision tree after seeding the initial population (compare with no seeding in Figure \ref{detection}c), and Figure \ref{seeding}b shows the performance of random forest after seeding the initial population (compare with no seeding in Figure \ref{detection}e), showing attacks with higher damage and low detection probability. These results indicate seeding could significantly reduce the duration of experiments, and more rapidly identify those attacks that are likely to evade detection (raise alarms) and, at the same time cause some economic loss. However, more experiments are required to understand the full benefits and weaknesses of the variety of strategies for this approach (e.g. such as the number of seeds used), as seeding can reduce the diversity of the population, and so fail to explore other feasible region in the attack space.

One of the weaknesses of the two objective optimisation against the detection methods was that the effort was not optimised, and we were not able to tell if attacks could be generated using less effort. To investigate this, if attackers could cause same damage using less effort, we carried out a 3-objective optimisation (maximise operating cost of the plant, minimise detection, minimise effort). Given the computational cost, only one detection method, AdaBoost was selected for further investigation. Table \ref{mupluslambdaresults} shows the performance of the NSGA-II and SPEA2 averaged over two runs, for 800 generations. 

\begin{table}[!htbp]
    \centering
    \begin{tabular}{c|c|c}
&   NSGA-II & SPEA2 \\
\hline
 Damage Range (\$) & 0-1633 & 0-1114 \\ 
 Hypervolume & 0.7140 & 0.5423 \\
 \hline
    \end{tabular}
    \caption{Performance of 3-objective attack generation against AdaBoost}
    \label{mupluslambdaresults}
\end{table}

\begin{figure} [!ht]
\centering
    \subcaptionbox{NSGA-II}{%
    \includegraphics[width=.47\linewidth]{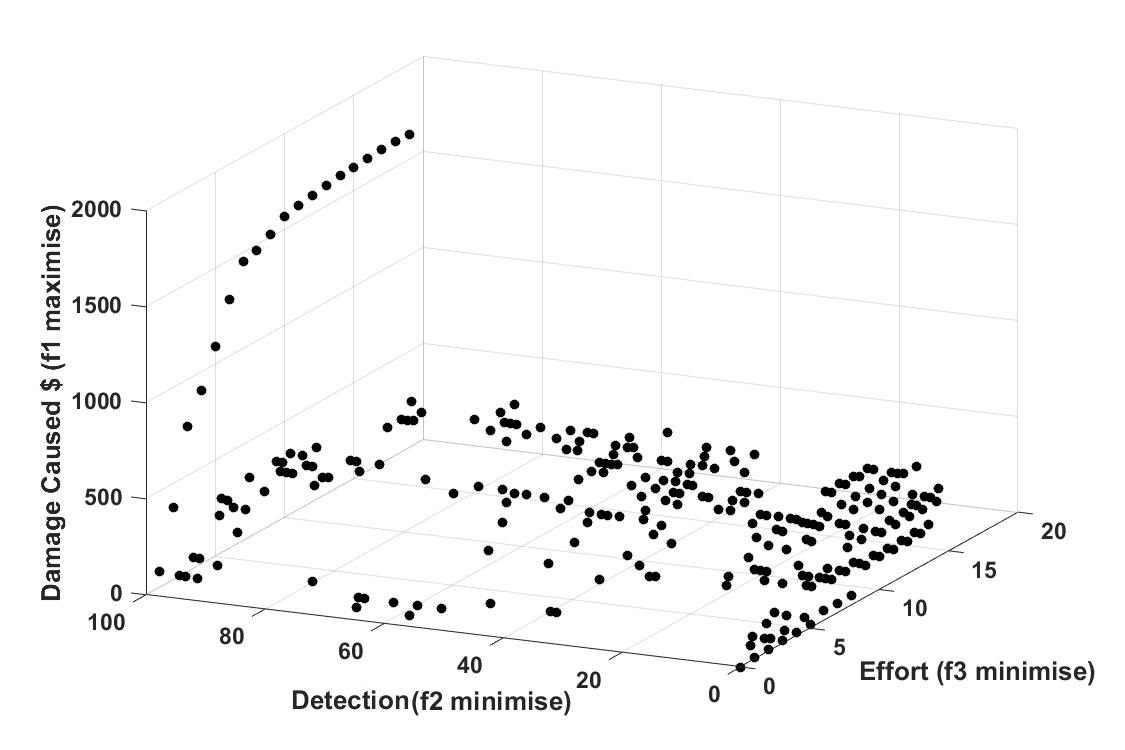}}\quad
  \subcaptionbox{SPEA2}{%
    \includegraphics[width=.47\linewidth]{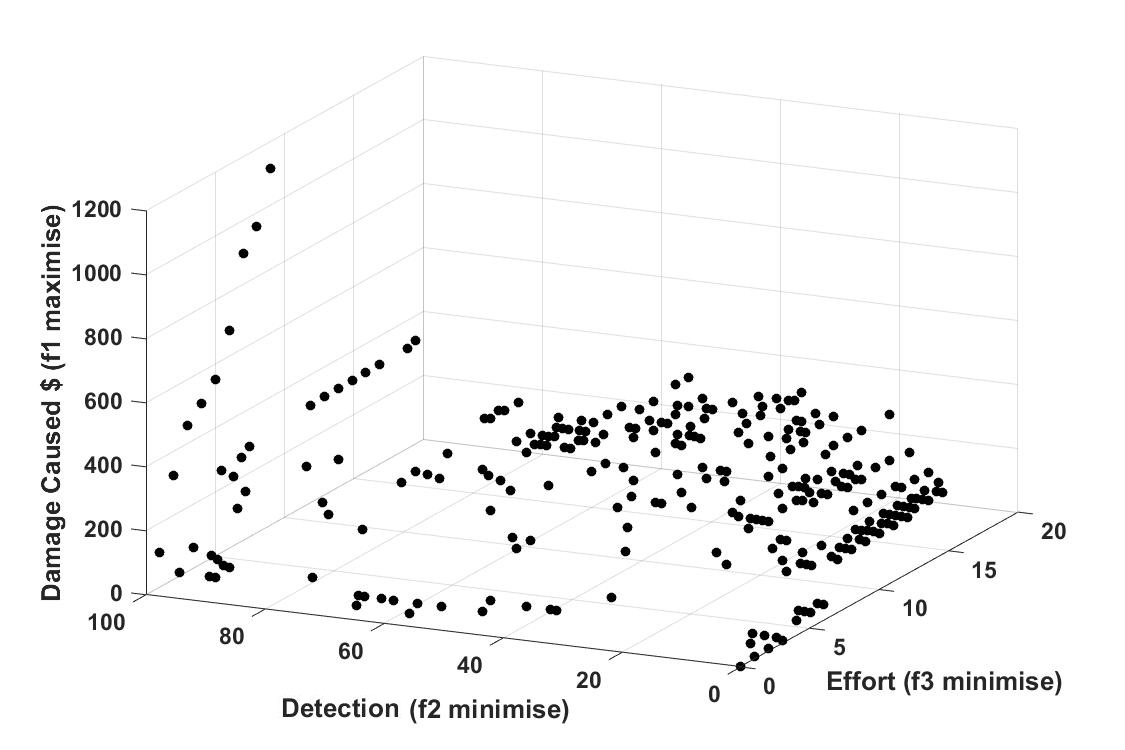}}
    \caption{Obtained Pareto front for 3-objective attack generation against AdaBoost}
    \label{paretodectionaverages}
\end{figure}

Both runs required more time to converge, but NSGA-II appears to search a wider region using 3-objectives compared to SPEA2. NSGA-II showed a better spread of attacks over the search space, and a better value for hypervolume. The best attacks are those with lower detection probability, and these tend to lie in the lower regions of the detection which cause damage less than \$200. As shown in Figure \ref{paretodectionaverages}, NSGA-II finds attacks with higher damage, damage over \$1500, but, SPEA2 tends to generate more attacks in the lower regions where there is a better probability of avoiding detection.

A more reliable comparison requires further experiments, which we leave as part of future work; however, overall, both algorithms found attacks that avoid detection and cause some harm. Assuming the detection probability threshold is less than 5\%, the highest cost attack NSGA-II found is \textdollar266.92, attacking 12 sensors and actuators, with a detection probability of 3\%, whereas SPEA2 found a slightly worse attack, \textdollar179.72 with effort 16 and detection probability of 4\%. If the intention is to use the smallest effort and cause maximum damage, then the optimal attack strategy produced using NSGA-II is an attack that costs \textdollar179.72 using effort 6. SPEA2 found a similar attack using effort 6, at a cost of \textdollar170.69. The attacks generated using less effort were either detected or caused very little economic damage, that is $\leq\textdollar50.00$. 

\section{Application of results}
The EMO approach developed in this paper can be employed to test the vulnerabilities of cyber-physical systems with a broad attack surface, where  attackers often target process measurements and manipulated variables, as these values can impact the process directly, and cause physical damage. 

Results obtained provide a rich set of attacks that can be analysed and help plant operators identify the most vulnerable combinations of sensors and actuators. Table \ref{tab2:vulnerableshut} shows a small subset of vulnerable combinations of the sensors and actuators that could shut down the TE plant in under 17 minutes. For example, carrying out a single attack on XMEAS 8 (reactor level) was able to shut down the plant in over 2.8 hours, and attacking XMEAS 11 (separator temperature) avoided shutting down the plant for all types of attacks, as shown in Table \ref{tablexmeas}. Results show that attacking both of these sensors simultaneously could shut down the plant in 14.8 minutes. These results also showed that the plant is less resilient to attacks on process measurements (sensors), and, if an adversary wants to bring down the plant in a very short period of time such as less than 10 minutes, attacking sensors is more likely to cause this to happen than attacking manipulated variables. This is, possibly because the selected attack parameters for actuator signals (based on the observed signals), were not as effective as attacking process measurements. Similarly, DoS attacks were slower and less successful against manipulated signals. Future research will investigate this further, focusing more on the attack parameters and timing of the attacks. 

\begin{table}[h]
\centering
    \begin{tabular}{| l | l |}
    \hline
   Vulnerable Combinations & SDT (min)  \\ \hline
XMEAS4,XMEAS8,XMEAS11, & 8.9  \\ 
XMEAS17,XMV6 & \\ \hline
XMEAS4,XMEAS8,XMEAS11, & 9.6 \\ 
XMEAS31 & \\ \hline
XMEAS3,XMEAS4,XMEAS8,& 10.2  \\
XMEAS11 & \\ \hline
XMEAS8,XMEAS9,XMEAS11 & 10.8 \\ \hline
XMEAS8,XMEAS11,XMV6  & 11.4 \\ \hline
XMEAS8,XMEAS11,XMV10  & 12.0  \\ \hline 
XMEAS5,XMEAS8,XMEAS11 & 13.8  \\ \hline
XMEAS9,XMV7,XMV11 & 14.4  \\ \hline
XMV7,XMV10,XMV11 & 14.5 \\ \hline
XMEAS4,XMEAS7,XMEAS17 & 14.7 \\ \hline
XMEAS8,XMEAS11 & 14.8 \\ \hline
XMV9, XMV10 & 16.2 \\ \hline
XMEAS9,XMV11 & 16.8 \\ \hline

    \end{tabular}
    \caption{A selection of vulnerable combinations that could bring the plant down under 17 minutes}
        \label{tab2:vulnerableshut}
\end{table}

Similarly, we found a wide variety of combined attacks that are capable of increasing the operating cost. Most of these attacks involved carrying out an $integrity_{max}$ attack on XMEAS 7 (reactor pressure), which means sending higher values than expected. When the controller receives these values, it attempts to reduce the pressure by opening the purge valve, and this causes raw materials to be lost in the purge stream. This in turn increases the operating cost. Carrying out a single attack on XMEAS 7 increases the operating cost to \$24,479 but, as the results indicate, more damage can be inflicted using the right combinations of sensors and actuators.  

Obtained results show EMO algorithms can be used to generate attacks to identify  vulnerable parts of a system, and this knowledge can be used to design systems that are more secure and resilient to cyber attacks. One way to achieve this is to consider the vulnerable combinations when designing network segmentation. The \textit{zone and conduit model} is a framework for network segmentation to manage security threats for industrial automation and control systems, recommended as part of the standard such as ISA/IEC 62443 \cite{ISA62443}. Zones are defined as a group of logical or physical assets sharing common security requirements, and conduits are the paths of communication between the zones. Using the knowledge from EMO, vulnerable combinations of sensors and actuators that cause the plant to shut down or increase the operating cost of the plant can be aggregated in different zones to build a more secure network. 

As we were able to generate a large number of attacks that were not detected by all detection methods, this approach can be also used as a tool to test and develop better detection methods. However, generating attacks that evade detection, and at the same time cause some significant economic damage on a system like TE process is a challenging task, since it requires attacks to be long in duration. Our results show that carrying out a successful attack requires knowledge of the system to ensure that the correct combination of sensors and actuators are attacked, and to avoid detection or the triggering of the safety system. The significant attacks generated against AdaBoost, decision tree and random forest classifiers involved attacking multiple components in the system to evade detection while causing economic damage. A na\"ive  attacker that randomly attacks multiple targets is unlikely to achieve similar damage, and is highly likely to be detected. The focus of our future work will involve investigating and designing more complex attacks that could learn the behaviour of the plant and the detection system, and generate attacks using this knowledge. 

Results obtained show that evolutionary multiobjective optimisation can be used successfully as an effective tool to model the behaviour of the adversary against a defence method, and illustrate the conflicts and associated trade-offs among common security objectives: attack impact, detection and effort required. Using the results obtained from such an analysis, security engineers can take measures to understand and eliminate system vulnerabilities before they are exploited by malicious actors.

\section{Conclusions and future work}
This paper demonstrates a novel and important application of evolutionary multiobjective optimisation for security of industrial control systems, and more general cyber-physical systems. Using a simulation of a complex and realistic plant, of the sort used routinely in factories and plants, it is possible to discover (and remove) vulnerabilities through an automatic process. The threat to such systems is both realistic and of critical importance, and there is no comparable mechanism that is robust, efficient and that allows the identification of vulnerable combinations of sensors and actuators. The knowledge gained from our work can be used in number of ways. The first is in determining the criticality of security decisions such as patching, carrying out risk assessment and designing resilient network segments. Secondly, control engineers can use the insight gained from this work to analyse the implications of security on process control, and design security-aware control algorithms. The attacks generated against the detection methods show our approach can also be used as a tool to find the vulnerabilities in the detection before these are exploited by the adversaries. 

The existing implementation of the TE model in MATLAB was too slow to be ideal for stochastic population-based optimisation algorithms like evolutionary algorithms and, therefore our experiments were limited by this. This was particularly a problem when generating attacks against the detection as evolution required more time to converge properly. Future work will involve optimising the code to improve the performance, and enable us to carry out attacks with better primitives. Planned future work will involve using the approach to design more advanced attacks against detection methods. TE model was used to illustrate the approach, however the approach is agnostic, and future work will focus on demonstrating this on other cyber-physical systems using more advanced attacks. The unavailability of accessible benchmark methods is a major obstacle for ICS security research. We are currently working on building cyber-physical testbeds, and we hope to test our approach on a different type of industrial process with physical and network components. Finally, will also investigate how other EMO algorithms, in particularly those based on aggregation-based and indicator-based algorithms to see if the obtained results can be further improved. 

\section{Acknowledgements}
This work was conducted partly under UK EPSRC Grant No: {EP/G037264/1} as part of University College London's Security Science Doctoral Training Centre, and partly under UK EPSRC Grant No: {EP/N023234/1} as part of PETRAS Internet of Things Research Hub.

\bibliographystyle{unsrt}  
\bibliography{template}  

\begin{thebibliography}{10}

\bibitem{PATTON2007280}
R.J. Patton, C.~Kambhampati, A.~Casavola, P.~Zhang, S.~Ding, and D.~Sauter.
\newblock {A Generic Strategy for Fault-Tolerance in Control Systems
  Distributed Over a Network}.
\newblock {\em European Journal of Control}, 13(2):280 -- 296, 2007.

\bibitem{evraz}
{Cyber attack shuts down Evraz IT systems across North America, but company
  says no data compromised}.
\newblock
  \url{https://www.cbc.ca/news/canada/saskatchewan/evraz-regina-shut-down-ransomware-attack-1.5487017}.
\newblock Last Accessed: 11/03/2020.

\bibitem{crashoverride}
{CRASHOVERRIDE Analyzing the Threat to Electric Grid Operations Version
  2.20170613}.
\newblock \url{https://dragos.com/wp-content/uploads/CrashOverride-01.pdf},
  June 2017.
\newblock Last Accessed: 04/05/2020.

\bibitem{german:steel}
{Die Lage der IT-Sicherheit in Deutschland 2014}.
\newblock Technical report, Bundesamt für Sicherheit in der
  Informationstechnik (BSI), Nov 2014.

\bibitem{duqu}
{W32.Duqu: The Precursor to the Next Stuxnet (Version 1.4)}.
\newblock Technical report, Symantec Security Response, Nov 2011.

\bibitem{havex}
{Havex Hunts For ICS/SCADA Systems}.
\newblock \url{https://www.f-secure.com/weblog/archives/00002718.html}.
\newblock Accessed: 02/05/2020.

\bibitem{Stuxnet}
{W32.Stuxnet Dossier (Version 1.4)}.
\newblock Technical report, Symantec Security Response, Feb 2011.

\bibitem{10.1145/1966913.1966959}
Alvaro~A. C\'{a}rdenas, Saurabh Amin, Zong-Syun Lin, Yu-Lun Huang, Chi-Yen
  Huang, and Shankar Sastry.
\newblock {Attacks against Process Control Systems: Risk Assessment, Detection,
  and Response}.
\newblock In {\em Proceedings of the 6th ACM Symposium on Information, Computer
  and Communications Security}, ASIACCS ’11, page 355–366, New York, NY,
  USA, 2011. Association for Computing Machinery.

\bibitem{HUANG200973}
Yu-Lun Huang, Alvaro~A. Cárdenas, Saurabh Amin, Zong-Syun Lin, Hsin-Yi Tsai,
  and Shankar Sastry.
\newblock Understanding the physical and economic consequences of attacks on
  control systems.
\newblock {\em International Journal of Critical Infrastructure Protection},
  2(3):73 -- 83, 2009.

\bibitem{Krotofil2013}
Marina Krotofil and Alvaro~A. C{\'a}rdenas.
\newblock {\em {Resilience of Process Control Systems to Cyber-Physical
  Attacks}}, pages 166--182.
\newblock Springer Berlin Heidelberg, Berlin, Heidelberg, 2013.

\bibitem{Genge12impactof}
B.~Genge, C.~Siaterlis, M.~Hohenadel, Béla Genge, Christos Siaterlis, and Marc
  Hohenadel.
\newblock Impact of network infrastructure parameters to the effectiveness of
  cyber attacks against industrial control systems.
\newblock {\em INTERNATIONAL JOURNAL OF COMPUTERS, COMMUNICATIONS \& CONTROL},
  page 673, 2012.

\bibitem{WANG201824}
Wei Wang, Antonio Cammi, Francesco~[Di Maio], Stefano Lorenzi, and Enrico Zio.
\newblock A monte carlo-based exploration framework for identifying components
  vulnerable to cyber threats in nuclear power plants.
\newblock {\em Reliability Engineering \& System Safety}, 175:24 -- 37, 2018.

\bibitem{10.1007/978-3-642-45330-4_15}
Antonio Di~Pietro, Chiara Foglietta, Simone Palmieri, and Stefano Panzieri.
\newblock {Assessing the Impact of Cyber Attacks on Interdependent Physical
  Systems}.
\newblock In Jonathan Butts and Sujeet Shenoi, editors, {\em Critical
  Infrastructure Protection VII}, pages 215--227, Berlin, Heidelberg, 2013.
  Springer Berlin Heidelberg.

\bibitem{8270567}
K.~{Huang}, C.~{Zhou}, Y.~{Tian}, S.~{Yang}, and Y.~{Qin}.
\newblock {Assessing the Physical Impact of Cyberattacks on Industrial
  Cyber-Physical Systems}.
\newblock {\em IEEE Transactions on Industrial Electronics}, 65(10):8153--8162,
  2018.

\bibitem{Li04usinggenetic}
Wei Li.
\newblock Using genetic algorithm for network intrusion detection.
\newblock In {\em In Proceedings of the United States Department of Energy
  Cyber Security Group 2004 Training Conference}, pages 24--27, 2004.

\bibitem{1460505}
T.~Xia, G.~Qu, S.~Hariri, and M.~Yousif.
\newblock An efficient network intrusion detection method based on information
  theory and genetic algorithm.
\newblock In {\em PCCC 2005. 24th IEEE International Performance, Computing,
  and Communications Conference, 2005.}, pages 11--17, April 2005.

\bibitem{5949394}
T.~Vollmer, J.~Alves-Foss, and M.~Manic.
\newblock Autonomous rule creation for intrusion detection.
\newblock In {\em 2011 IEEE Symposium on Computational Intelligence in Cyber
  Security (CICS)}, April 2011.

\bibitem{gaids}
AA~Ojugo, AO~Eboka, O~Okonta, RE~Yoro, and FO~Aghware.
\newblock {Genetic algorithm rule-based intrusion detection system (GAIDS)}.
\newblock {\em Journal of Emerging Trends in Computing and Information
  Sciences}, 2012.

\bibitem{DBLP:journals/corr/abs-1204-1336}
Mohammad~Sazzadul Hoque, Md.~Abdul Mukit, and Md. Abu~Naser Bikas.
\newblock {An Implementation of Intrusion Detection System Using Genetic
  Algorithm}.
\newblock {\em CoRR}, abs/1204.1336, 2012.

\bibitem{Diaz-gomez05improvedoff-line}
Pedro~A. Diaz-gomez, Ingenieria~De Sistemas, and Dean~F. Hougen.
\newblock Improved off-line intrusion detection using a genetic algorithm.
\newblock In {\em in Proceedings of the Seventh International Conference on
  Enterprise Information Systems}, 2005.

\bibitem{Goyal2008}
Anup Goyal and Chetan Kumar.
\newblock {GA-NIDS: A Genetic Algorithm based Network Intrusion Detection
  System}.
\newblock In {\em ElectricalEngineering \& Computer Science, North West
  University, Technical Report}, 2008.

\bibitem{COIN:COIN247}
Wei Lu and Issa Traore.
\newblock {Detecting New Forms of Network Intrusion Using Genetic Programming}.
\newblock {\em Computational Intelligence}, 20(3):475--494, 2004.

\bibitem{5718700}
S.~Pastrana, A.~Orfila, and A.~Ribagorda.
\newblock {A Functional Framework to Evade Network IDS}.
\newblock In {\em 2011 44th Hawaii International Conference on System
  Sciences}, pages 1--10, Jan 2011.

\bibitem{Mrugala:2016:GECCOcomp}
Kinga Mrugala, Nilufer Tuptuk, and Stephen Hailes.
\newblock {Evolving Attackers against Wireless Sensor Networks}.
\newblock In {\em GECCO 2016 Companion Volume}, page pp306, Denver, USA, 20-24
  July 2016. ACM.

\bibitem{7979715}
K.~Mrugala, N.~Tuptuk, and S.~Hailes.
\newblock {Evolving attackers against wireless sensor networks using genetic
  programming}.
\newblock {\em IET Wireless Sensor Systems}, 7(4):113--122, 2017.

\bibitem{Kayacik:2006:EBO:1143997.1144271}
Hilmi~G\"{u}ne\c{s} Kayacik, Malcolm Heywood, and Nur Zincir-Heywood.
\newblock {On Evolving Buffer Overflow Attacks Using Genetic Programming}.
\newblock In {\em Proceedings of the 8th Annual Conference on Genetic and
  Evolutionary Computation}, GECCO '06, pages 1667--1674, New York, NY, USA,
  2006. ACM.

\bibitem{John:2014:EBM:2598394.2605437}
David~J. John, Robert~W. Smith, William~H. Turkett, Daniel~A. Ca\~{n}as, and
  Errin~W. Fulp.
\newblock {Evolutionary Based Moving Target Cyber Defense}.
\newblock In {\em Proceedings of the Companion Publication of the 2014 Annual
  Conference on Genetic and Evolutionary Computation}, GECCO Comp '14, pages
  1261--1268, New York, NY, USA, 2014. ACM.

\bibitem{Dewri:2007:OSH:1315245.1315272}
Rinku Dewri, Nayot Poolsappasit, Indrajit Ray, and Darrell Whitley.
\newblock {Optimal Security Hardening Using Multi-objective Optimization on
  Attack Tree Models of Networks}.
\newblock In {\em Proceedings of the 14th ACM Conference on Computer and
  Communications Security}, CCS '07, pages 204--213, New York, NY, USA, 2007.
  ACM.

\bibitem{Garcia:2017:ICA:3067695.3076081}
Dennis Garcia, Anthony~Erb Lugo, Erik Hemberg, and Una-May O'Reilly.
\newblock {Investigating Coevolutionary Archive Based Genetic Algorithms on
  Cyber Defense Networks}.
\newblock In {\em Proceedings of the Genetic and Evolutionary Computation
  Conference Companion}, GECCO '17, pages 1455--1462, New York, NY, USA, 2017.
  ACM.

\bibitem{10.1145/3205651.3208287}
Erik Hemberg, Joseph~R. Zipkin, Richard~W. Skowyra, Neal Wagner, and Una-May
  O’Reilly.
\newblock {Adversarial Co-Evolution of Attack and Defense in a Segmented
  Computer Network Environment}.
\newblock In {\em Proceedings of the Genetic and Evolutionary Computation
  Conference Companion}, GECCO ’18, page 1648–1655, New York, NY, USA,
  2018. Association for Computing Machinery.

\bibitem{Rush:2015:CAN:2739482.2768429}
George Rush, Daniel~R. Tauritz, and Alexander~D. Kent.
\newblock {Coevolutionary Agent-based Network Defense Lightweight Event System
  (CANDLES)}.
\newblock In {\em Proceedings of the Companion Publication of the 2015 Annual
  Conference on Genetic and Evolutionary Computation}, GECCO Companion '15,
  pages 859--866, New York, NY, USA, 2015. ACM.

\bibitem{4290990}
T.~Service, D.~Tauritz, and W.~Siever.
\newblock {Infrastructure Hardening: A Competitive Coevolutionary Methodology
  Inspired by Neo-Darwinian Arms Races}.
\newblock In {\em Computer Software and Applications Conference, 2007. COMPSAC
  2007. 31st Annual International}, volume~1, pages 101--104, July 2007.

\bibitem{doi:10.1142/S1793005709001416}
Travis Service and Daniel Tauritz.
\newblock {Increasing Infrastructure Resilience through Competitive
  Coevolution}.
\newblock {\em New Mathematics and Natural Computation}, 05(02):441--457, 2009.

\bibitem{5679047}
J.~Decraene, M.~Chandramohan, M.~Y.~H. Low, and C.~S. Choo.
\newblock {Evolvable simulations applied to Automated Red Teaming: A
  preliminary study}.
\newblock In {\em Simulation Conference (WSC), Proceedings of the 2010 Winter},
  pages 1444--1455, Dec 2010.

\bibitem{8506545}
R.~Bronfman-Nadas, N.~Zincir-Heywood, and J.~T. Jacobs.
\newblock {An Artificial Arms Race: Could it Improve Mobile Malware Detectors?}
\newblock In {\em 2018 Network Traffic Measurement and Analysis Conference
  (TMA)}, pages 1--8, June 2018.

\bibitem{DOWNS1993245}
J.J. Downs and E.F. Vogel.
\newblock A plant-wide industrial process control problem.
\newblock {\em Computers \& Chemical Engineering}, 17(3):245 -- 255, 1993.

\bibitem{ie000586y}
Truls Larsson, Kristin Hestetun, Espen Hovland, and Sigurd Skogestad.
\newblock {Self-Optimizing Control of a Large-Scale Plant: The Tennessee
  Eastman Process}.
\newblock {\em Industrial \& Engineering Chemistry Research},
  40(22):4889--4901, 2001.

\bibitem{Ricker15}
N.~Lawrence Ricker.
\newblock {Tennessee Eastman Challenge Archive}.
\newblock
  \url{http://depts.washington.edu/control/LARRY/TE/download.html#Multiloop},
  Dec 1998.

\bibitem{IsakovKrotofil}
A.~Isakov and M.~Krotofil.
\newblock {Damn Vulnerable Chemical Process - Tennessee Eastman}.
\newblock \url{https://github.com/satejnik/DVCP-TE}, 2015.

\bibitem{7360024}
N.~Riquelme, C.~Von Lücken, and B.~Baran.
\newblock Performance metrics in multi-objective optimization.
\newblock In {\em 2015 Latin American Computing Conference (CLEI)}, pages
  1--11, Oct 2015.

\bibitem{FIELDER201613}
Andrew Fielder, Emmanouil Panaousis, Pasquale Malacaria, Chris Hankin, and
  Fabrizio Smeraldi.
\newblock {Decision support approaches for cyber security investment}.
\newblock {\em Decision Support Systems}, 86:13 -- 23, 2016.

\bibitem{RICKER1995949}
N.L. Ricker.
\newblock {Optimal steady-state operation of the Tennessee Eastman challenge
  process}.
\newblock {\em Computers \& Chemical Engineering}, 19(9):949 -- 959, 1995.

\bibitem{Deb2000}
Kalyanmoy Deb, Samir Agrawal, Amrit Pratap, and T.~Meyarivan.
\newblock {\em {A Fast Elitist Non-dominated Sorting Genetic Algorithm for
  Multi-objective Optimization: NSGA-II}}, pages 849--858.
\newblock Springer Berlin Heidelberg, Berlin, Heidelberg, 2000.

\bibitem{996017}
K.~Deb, A.~Pratap, S.~Agarwal, and T.~Meyarivan.
\newblock {A fast and elitist multiobjective genetic algorithm: NSGA-II}.
\newblock {\em IEEE Transactions on Evolutionary Computation}, 6(2):182--197,
  April 2002.

\bibitem{Zitzler01spea2:improving}
Eckart Zitzler, Marco Laumanns, and Lothar Thiele.
\newblock {SPEA2: Improving the Strength Pareto Evolutionary Algorithm}.
\newblock Technical report, Swiss Federal Institute of Technology (ETH), 2001.

\bibitem{Fortin:2012:DEA:2503308.2503311}
F{\'e}lix-Antoine Fortin, Fran\c{c}ois-Michel De~Rainville,
  Marc-Andr{\'e}~Gardner Gardner, Marc Parizeau, and Christian Gagn{\'e}.
\newblock {DEAP: Evolutionary Algorithms Made Easy}.
\newblock {\em J. Mach. Learn. Res.}, 13(1):2171--2175, July 2012.

\bibitem{Zitzler:2000:CME:1108872.1108876}
Eckart Zitzler, Kalyanmoy Deb, and Lothar Thiele.
\newblock {Comparison of Multiobjective Evolutionary Algorithms: Empirical
  Results}.
\newblock {\em Evol. Comput.}, 8(2):173--195, June 2000.

\bibitem{10.1007/BFb0056872}
Eckart Zitzler and Lothar Thiele.
\newblock {Multiobjective optimization using evolutionary algorithms - A
  comparative case study}.
\newblock In Agoston~E. Eiben, Thomas B{\"a}ck, Marc Schoenauer, and Hans-Paul
  Schwefel, editors, {\em Parallel Problem Solving from Nature - PPSN V}, pages
  292--301, Berlin, Heidelberg, 1998. Springer Berlin Heidelberg.

\bibitem{870296}
D.~A. {Van Veldhuizen} and G.~B. {Lamont}.
\newblock On measuring multiobjective evolutionary algorithm performance.
\newblock In {\em Proceedings of the 2000 Congress on Evolutionary Computation.
  CEC00 (Cat. No.00TH8512)}, volume~1, pages 204--211 vol.1, July 2000.

\bibitem{FRIEDRICH2015223}
Tobias Friedrich and Markus Wagner.
\newblock {Seeding the initial population of multi-objective evolutionary
  algorithms: A computational study}.
\newblock {\em Applied Soft Computing}, 33:223 -- 230, 2015.

\bibitem{ISA62443}
{ISA99: Developing the Vital ISA/IEC 62443 Series of Standards on Industrial
  Automation and Control Systems (IACS) Security}, 2015.

\end{thebibliography}

\end{document}